\documentclass{ui-phdthesis}
\usepackage{multirow}
\usepackage{fancyheadings}
\usepackage[para]{threeparttable}
\usepackage{amsmath}
\usepackage{amssymb}
\usepackage{epsfig}
\usepackage{graphicx}
\usepackage[document]{ragged2e}
\usepackage{float}
\usepackage{here}
\usepackage{longtable}
\usepackage{footnote}
\usepackage{pdflscape}
\usepackage{titlesec}
\usepackage[font=small,labelfont=bf,justification=justified,format=plain]{caption}
\usepackage[colorlinks=true,linkcolor=blue,citecolor=blue,urlcolor=blue]{hyperref}
\usepackage{natbib}
\usepackage{booktabs}
\usepackage{changepage}
\usepackage{aas_macros}
\usepackage{wallpaper}
\usepackage{subfigure}
\usepackage{blindtext}
\usepackage{chngcntr}
\usepackage{sectsty}
\counterwithout{figure}{chapter}
\counterwithout{table}{chapter}
\counterwithin{figure}{section}
\counterwithin{table}{section}
\usepackage[font=small,labelfont=bf,
justification=justified,
format=plain]{caption}

\usepackage{hyperref}%

\usepackage{txfonts} 

\newcommand{\lya}{Ly$\alpha$}

\newcommand{\ci}{C\,{\sc i}}

\newcommand{\hi}{H\,{\sc i}}

\newcommand{\farcs}{\mbox{\ensuremath{.\!\!^{\prime\prime}}}}
\newcommand{\kms}{\,km\,s$^{-1}$}

\thesisauthor{Kasper Elm Heintz}
\thesistitle{Galaxies through Cosmic Time Illuminated by Gamma-Ray Bursts and Quasars}
\thesisshorttitle{ }
\thesismonth{July}
\thesisyear{2019}
\thesisdegree{Philosophiae Doctor}
\thesisfield{Physics}
\thesisschool{School of Engineering and Natural Sciences}
\thesisfaculty{Physical Sciences}
\thesisaddress{Dunhagi}
\thesispostalcode{107}
\thesistelephone{525-4000}
\thesisprinter{Háskólaprent} 
\thesiscommittee{\it Prof. P\'all Jakobsson (supervisor) \\ Assoc. Prof. Jes\'us Zavala \\ Prof. Emeritus Einar H. Guðmundsson}
\thesisopponents{\it Prof. J. Xavier Prochaska \\ Dr. Valentina D'Odorico}
\thesisISBN{978-9935-9473-3-8} 
\begin{document}
\makecovertitlepage
\makeinnertitlepage
\makesaurblad
\thesislists
\justify
\thesisabstract{TRUE}{In the early Universe, most of the cold neutral gas that will later form into individual stars and galaxies is practically invisible to us. These neutral gas reservoirs can, however, be illuminated by bright cosmic lightsources such as gamma-ray bursts (GRBs) and quasars. The aim of this thesis is to use these luminous objects as tools to study the environments of intervening or host galaxy absorption systems through cosmic time. 

\vspace{0.3cm}

{\bf Part I} of this thesis is dedicated to examining the gas, dust and metals in the immediate region surrounding GRBs. In Chapter~\ref{chap:140506a}, a comprehensive analysis of the host galaxy of GRB\,140506A is presented. The optical afterglow spectrum of this GRB showed a very peculiar shape, likely related to an unusual dust reddening curve that extinguished the intrinsic afterglow spectrum. We were able to better characterize this unusual afterglow extinction by subtracting the host galaxy spectrum from the first epoch afterglow spectrum and showed that it can be modelled by a very steep reddening law (but not a 2175\,\AA~dust bump as previously proposed). Since no such extinction component was observed for the average host galaxy spectral energy distribution (SED), we concluded that it was produced local to the GRB, and likely related to the interaction of the GRB with the surrounding circumburst dust. In Chapter~\ref{chap:highion}, the high-ionization metal lines in a sample of GRB afterglows observed with the VLT/X-shooter are examined. We argue that in particular the N\,{\sc v} absorption line feature is produced by the surrounding gas that has been highly ionized by the GRB. This argument is also based on the observed tentative corelation with the X-ray derived equivalent hydrogen column density, $N_{\rm H,X}$, that is also believed to originate from gas within $\sim 10$\,pc from the GRB. The highly-ionized N\,{\sc v} line transition therefore likely probes gas and metals in the circumburst environment of the GRB.

\vspace{0.2cm}

{\bf Part II} presents a search for and the study of cold and molecular gas in high-$z$ GRB host galaxy absorption systems. Chapter~\ref{chap:cisurvey} introduces a survey for neutral atomic-carbon (\ci), which is used as a tracer for molecular hydrogen (H$_2$) in the GRB host absorbers. Here, we also characterize the basic properties of the GRB {\ci} absorbers in terms of their gas, dust and metal content and compare them to {\ci}-bearing absorbers in quasar sightlines. We find that a higher H\,{\sc i} column density and metallicity is required for GRB host galaxies to contain cold and molecular gas. In Chapter~\ref{chap:cidust}, we specifically examine the dust properties of the GRB {\ci} absorbers and demonstrate that the characteristic 2175\,\AA~dust extinction feature is likely produced by dust particles associated with the {\ci}-bearing molecular cloud. Chapter~\ref{chap:h2physprop} provides a detailed analysis of the presence of different probes of the molecular gas-phase in the ISM of GRB host galaxies such as {\ci} and vibrationally-excited H$_2$. Here, the typical physical conditions of the H$_2$-bearing molecular clouds in GRB host galaxies are also derived.

\vspace{0.2cm}

{\bf Part III} focuses on using quasars to examine gas-rich intervening galaxies in the line of sight, with specific focus on absorption systems rich in dust and metals. It is now clear that several dust-reddened quasars have evaded identification in e.g. the extensive Sloan Digital Sky Survey (SDSS). To alleviate this dust bias we designed the new optical to near/mid-infrared selection criteria presented in Chapter~\ref{chap:kvrq}, specifically tailored to select dust-reddened quasars at $z>2$. In Chapter~\ref{chap:dustydla}, one of the dusty quasar absorption systems identified in this survey is presented. Here, we also characterize the properties of this absorber and compare the few cases of dusty and metal-rich absorption systems to the general population of absorbers observed in quasar sightlines. 
Finally, in Chapter~\ref{chap:gaiasel} we demonstrate that it is possible to define a complete and unbiased quasar selection using the astrometric data from the space-based {\it Gaia} mission. 

\vspace{0.3cm}

This thesis demonstrates the importance of observing large samples of GRB afterglows to 1) allow for statistical studies of the GRB phenomena itself and the associated host galaxy environments and 2) to obtain spectra of peculiar or unusual GRB afterglows, that is only observed rarely. In addition, it highlights that defining a complete and unbiased sample of quasars is vital to fully exploit the potential of quasars as probes of cosmic chemical evolution. In Chapter~\ref{chap:conc}, the general work of this thesis is concluded and some future prospects are provided.}

\thesisutdrattur{Megnið af hinu kalda, ójónaða efni í ungum alheimi, sem
síðar myndar stjörnur og vetrarbrautir, er nánast ósýnilegt
okkur. Þó geta björt fyrirbæri eins og gammablossar og dulstirni
lýst upp þetta dimma efni. Markmið ritgerðarinnar er að nota þessi
fyrirbæri sem nokkurs konar verkfæri til að fræðast um útgeimsefnið.

\vspace{0.3cm}

{\bf Í fyrsta hluta} ritgerðarinnar er gas, ryk og þungefni í nánasta
nágrenni gammablossa kannað til hlítar. Í kafla~\ref{chap:140506a} er kynnt
yfirgripsmikil greining á hýsilvetrarbraut blossans GRB\,140506A.
Litróf sýnilegra glæða blossans var mjög óvenjulegt, líklega
tengt sérstæðri geimroðnun. Okkur tókst að sýna fram á að þessi
sérkennilega ljósdeyfing ætti upptök nálægt blossanum og væri
afleiðing af víxlverkun blossans við sitt nánasta umhverfi. Í
kafla~\ref{chap:highion} er safn blossa skoðað sem rannsakað var með nýlegum
litrófsmæli á Very Large Telescope (VLT). Hér var áhersla lögð
á gleypilínur frá mikið jónuðu gasi. Við sýndum fram á að þessar
línur eru afleiðing blossans og notuðum auk þess röntgenmælingar
til að styrkja þær niðurstöður.

\vspace{0.2cm}

{\bf Í öðrum hluta} er lögð áhersla á rannsóknir á sameindagasi við
hátt rauðvik. Í \ref{chap:cisurvey}. kafla er hlutlaust kolefni (\ci) kortlagt
en það má nota sem nokkurs konar sporefni fyrir vetnissameindina
H$_2$. Grunneiginleikar viðkomandi gass voru rannsakaðir og
meginniðurstaðan var sú að hýsilvetrarbrautir gammablossa verða
að hafa hátt hlutfall þungefna til að geta innihaldið kalt
sameindagas. Í \ref{chap:cidust}. kafla er athyglinni beint að eiginleikum ryks
í {\ci} kerfum og í \ref{chap:h2physprop}. kafla er nákvæm greining á tilvist sporefna
sameindagass í miðgeimsefni hýsilvetrarbrauta gammablossa.

\vspace{0.2cm}

{\bf Í þriðja} og síðasta hluta eru dulstirnin í aðalhlutverki. Þau
eru hér notuð til að rannsaka kerfi sem eru rykrík og hafa hátt
hlutfall þungefna. Kynnt er ný aðferð til að finna slík kerfi
við hátt rauðvik sem fyrri athuganir hafa hingað til misst af.
Fjallað er um slíka valforsendu, sem byggir á sýnilegum og
innrauðum mælingum, í kafla~\ref{chap:kvrq}. Eitt slíkt kerfi er svo rannsakað
gaumgæfilega í kafla~\ref{chap:dustydla}. Að lokum eru gögn frá {\it Gaia} geimsjónaukanum notuð í kafla~\ref{chap:gaiasel} til að sýna fram á að hægt er að skilgreina óhlutdrægt safn dulstirna.

\vspace{0.3cm}

Þessi ritgerð sýnir fram á mikilvægi þess að skoða stór söfn
gammablossa til að 1) gera tölfræðilegar rannsóknir á þeim og
hýsilvetrarbrautum þeirra sem og að 2) auka möguleikana á því að
uppgötva sérkennilegt umhverfi sumra þeirra. Að auki þá undirstrika
þessar rannsóknir brýna þörf á að skilgreina óhlutdrægt safn dulstirna.
Þannig má nýta þau til að kortleggja aukningu þungefnis sem fall af
tíma í alheiminum.
}


%
%




\thesisacknowledgments{TRUE}{
\begin{flushright}
	\textit{"If I have seen further, it is by \\ standing on the shoulders of Giants"} \\
	- Isaac Newton, 1675\footnote{Modifications of this saying have been traced back to the 12th century, but the familiar English expression is from a letter to Robert Hooke from Isaac Newton, sent in 1675.}
\end{flushright}

\vspace{0.2cm}

\noindent I think this quote is excellent in realizing the importance of the previously laid out groundwork any future advances in science relies on. For me personally, it also refers to my supervisors and mentors throughout my time as a PhD student, without whom the work done in this thesis would not have been possible. 

First of all, I would like to thank my main supervisor Palli. You accepted me as a PhD student even though I had to spend the first half of my studies abroad in Copenhagen. You allowed me to pursue my childhood dream of studying astrophysics and for that I am deeply grateful. Then I would also like to thank Johan for accepting to be my unofficial supervisor in Copenhagen, and for always being there to discuss any problems I might have, scientific or otherwise. I am sure that I would not be where I am today without your help and guidance. I greatly appreciate the amount of freedom both of you have given me during my PhD studies. I am also thankful to both of you for including me in the GRB afterglow follow-up team as part of the XS-GRB/Stargate collaboration, which have been, and continue to be, very exciting. Related to this, I also owe a special thanks to Daniele for preparing me for the intense work that follows a GRB trigger and for inviting me to the ePESSTO team, which resulted in a great trip to the La Silla observatory in Chile together. Finally, I would also like to thank Palle for acting as a scientific mentor for me and for first showing me how astronomy is done outside of Copenhagen.

It has been a great privilege to be included in the GRB collaboration and the mature field of GRBs in general, guided by the foremost experts, and to work on this unique data set that has been collected throughout almost a decade as part of my PhD project. As part of this collaboration, I have had numerous experts on GRBs evaluate my work, whose knowledge has greatly expanded my own. Specifically, I would like to thank Darach for our great discussions on about almost everything, but in particular on the work done as part of our high-ion project and all of our ongoing projects. I would also like to thank C\'edric and Pasquier for enlightening discussions during our work on cold and molecular gas and in general eminent scientific guidance.

\newpage

I probably owe my greatest thanks to my now wife, Julie, for her endless support throughout my PhD studies and for agreeing (though reluctantly in the beginning) to move to Iceland with me so that I could pursue my dream. It has been quite an adventure with you and our son, Oliver. Finally, I would like to thank my mom and dad for always believing in me and supporting me through my entire life. You sparked my interest for physics and astronomy in the first place.

}
\thesisbody
\chapterfont{\centering}

\chapter{Introduction}

The topics covered in this thesis are roughly divided into three parts. The first part (Chapters~\ref{chap:140506a} and \ref{chap:highion}) is devoted to the study of the very local environment in which gamma-ray burst (GRB) progenitors reside. Chapters~\ref{chap:cisurvey} to \ref{chap:h2physprop} constitute the second part, focusing on the cold and molecular gas-phase of the interstellar medium (ISM) in GRB host galaxies. The work in these first two parts is based on a combination of understanding a single, extraordinary burst and the resulting implications for the global GRB population in addition to a detailed characterization of a larger sample of GRB afterglows allowing for statistical analyses of a more homogenous selected sample. Both approaches are valuable to gauge the true underlying characteristics of the GRB host galaxy population and emphasizes the need for continuous follow-up of GRBs to identify single, peculiar cases and building large samples to draw general conclusions about the GRB host galaxy environments. 

The last part (Chapters~\ref{chap:kvrq} through \ref{chap:gaiasel}) is devoted to the study of quasars and intervening absorption systems, and in particular how current quasar selection techniques are subject to severe biases. Specifically, quasars with foreground dusty and metal-rich absorbers are evasive to classical quasar surveys, so in this part it is demonstrated how the selection of these bright cosmic sources can be modified to alleviate this bias. Below follows a brief, general introduction to the nature of GRBs and quasars, with particular focus on their use as cosmic probes. In Sect.~\ref{sec:dlas}, a more comprehensive introduction will be given to the study of absorption-line systems in GRB and quasar spectra.

\section{The nature of GRBs and quasars}

The central engine powering the light emitted from GRBs and quasars are believed to be produced by roughly the same mechanism, though on vastly different scales. Both phenomena are related to the accretion of gas onto black holes, extracting angular momentum and potential energy into electromagnetic radiation via the Blandford-Znajek mechanism \citep{Blandford77}. In this process, the strong induced magnetic fields will also form relativistic jets launched along the rotation axis of the accreting gas. In the case of (long-duration) GRBs, a stellar-mass black hole is formed by the core collapse of a single massive star, whereas the super-massive black holes observed to reside in the centre of most galaxies can reach masses of $M = 10^5 - 10^9\,M_{\odot}$ via merging of smaller black holes and accretion of matter, and are able to explain the basic properties of quasars \citep{Salpeter64}. 

While the story of how GRBs were first discovered and the following attempts to piece together the puzzle of these cosmic explosions is interesting on its own (see every PhD theses on GRBs ever), I will in the following sections only outline the basic nature and properties of these energetic $\gamma$-ray flashes and subsequent emission components. I will also only briefly introduce the phenomenom of quasars as well, but still include a short description on how these were first discovered and subsequently identified in Sect.~\ref{sec:qsosel}, since it is important for the later topics of this thesis. The overview provided in this section thus only highlights the relevant characteristics of GRBs and quasars and their application as cosmic tools and is not meant as a comprehensive introduction to either of the two phenomena.

\subsection{GRB optical afterglows} \label{ssec:grbaft}

With the first collection of GRBs observed by the Burst And Transient Source Experiment (BATSE) onboard the Compton Gamma Ray Observatory (CGRO), it was possible to constrain two fundamental properties of the overall GRB population. First, the angular distribution was observed to be isotropic on the sky, favoring a cosmological, instead of the previously proposed Galactic, origin \citep{Meegan92}. Second, the burst durations appeared to follow an apparent bimodal distribution (typically quantified as $t_{90}$, the time in which 90\% of the source photons are collected), with a division around 2 seconds \citep{Kouveliotou93}. While not being clear at the time, it is now believed that the \lq short\rq~($t_{90} < 2$\,s) and \lq long\rq~($t_{90} > 2$\,s) GRBs are actually produced by two distinct physical phenomena. 
Later, several more characteristics have been used to divide the two sub-populations since the typically used observables show a large degree of overlap \citep[e.g.][]{Lien16}. For instance, short GRBs typically have a higher fraction of the radiation emitted at higher energies, which is often referred to as having \lq harder\rq~spectra. On the other hand, the total output energies are typically lower than observed for long GRBs resulting in significantly fainter X-ray and optical afterglows \citep{Berger14,DAvanzo15}. 

The canonical prescription for long GRBs is the so-called \lq collapsar\rq~model \citep{Woosley93,MacFadyen99}, where a black hole is formed by the core collapse of a massive ($M_{\star} \gtrsim 40\,M_{\odot}$) Wolf-Rayet star. Short GRBs are instead believed to be produced by the merger of binary neutron stars or a neutron star and a stellar-mass black hole \citep[e.g.][]{Narayan92}. While the two GRB progenitor systems are physically distinct, the initial emission mechanisms are expected to be roughly the same.
The energy output of both types of GRBs are separated into two phases, the prompt and afterglow emission. The prompt emission is characterized by an intense $\gamma$-ray flux and is produced by ultrarelativistic flows that gets dissipated in internal collisions along the jet direction. When the jet of $\gamma$-rays then collides with the surrounding medium it gets slowed down by external shocks and emits synchrotron radiation which is observed as the X-ray to optical and radio afterglow \citep{Piran04}. The spectral shape of the afterglow typically follows a simple, temporally varying power-law as $F_{\lambda} = \lambda^{-\beta}t^{-\alpha}$ \citep{Sari98}. 

The still on-going {\it Swift} mission \citep{Gehrels04} has in the more recent years significantly contributed to our current understanding of GRBs \citep[see e.g.][]{Zhang07,Gehrels09}. The typical precision of the on-sky localization of the X-ray afterglow is confined within an error circle of around 5 arcsec, which has substantially increased the number of detected optical afterglows. This allows for a better understanding of the afterglow itself, but also of the associated transients that provide imperative clues to the GRB progenitors.

\subsection{Late-stage emission components associated with GRBs}

\begin{figure}[!t]
	\centering
	\includegraphics[width=12cm]{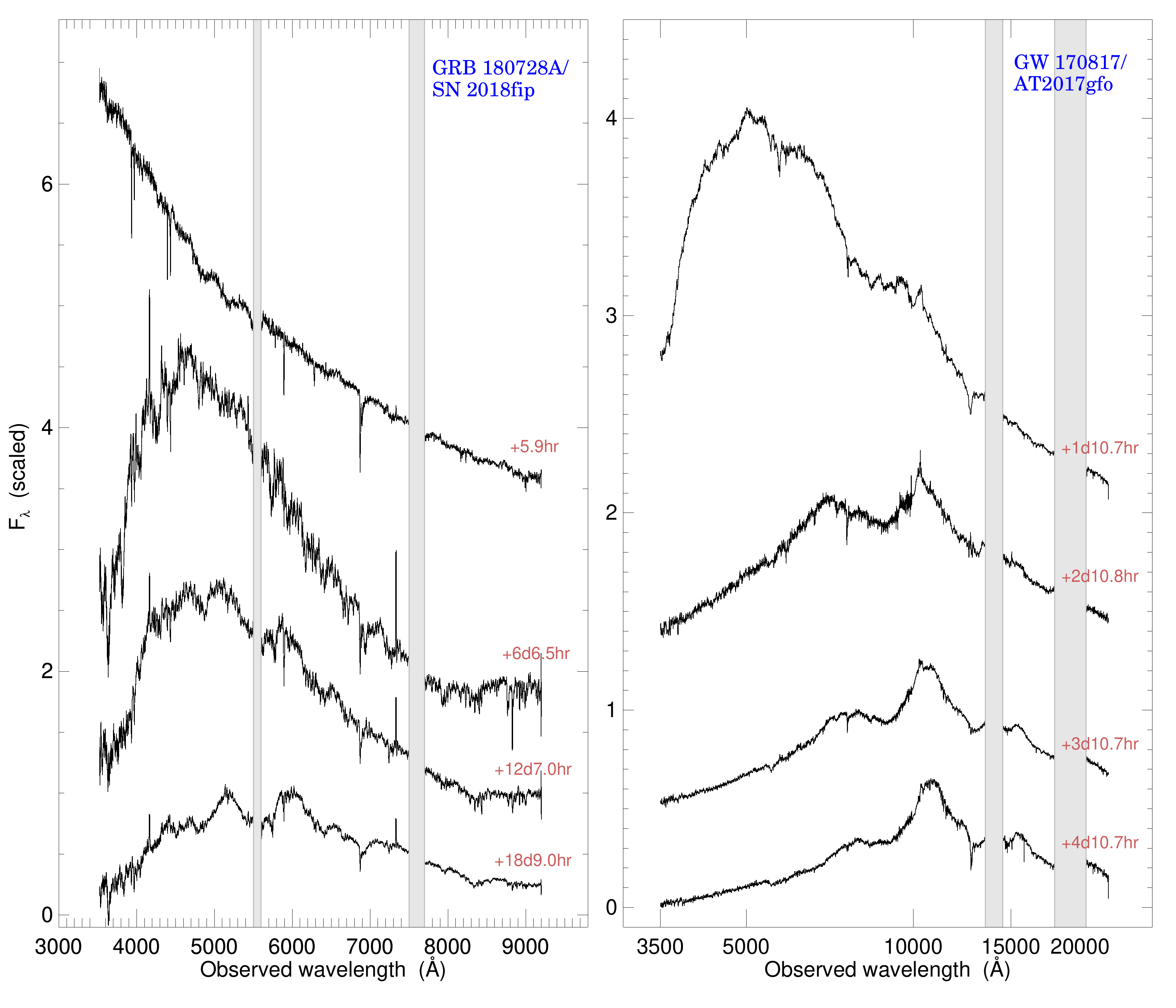}
	\caption{Spectral evolution of the long-duration GRB\,180728A (left panel) and the short-duration GRB and gravitational wave counterpart GW\,170817/AT\,2017gfo (right panel). The left panel shows the evolution of typical long-duration GRBs that transitions into broad-lined type Ic supernova over the course of a few weeks. The right panel shows the first four VLT/X-shooter spectra obtained of the kilonova AT\,2017gfo accompanying the gravitational wave-detected neutron star merger GW\,170817, that first shows thermal emission from a $T\approx 5000$\,K plasma that quickly cools and transitions into being dominated by wide spectral features in the near-infrared. The temporal evolution happens on much shorter time-scales compared to typical long-duration GRB-SNe.
	Both spectral sequences have been artifically offset along the $y$-axis for a more clear comparison.}
	\label{fig:grbsngwkn} 
\end{figure}

Shortly after the discovery of the first afterglow from a long GRB, conclusive evidence for the stellar origin came from the detection of an associated broad-lined type Ic supernova \citep[SN Ic-BL;][]{Galama98,Woosley99}. This particular type of SN show highly broadened spectral features with an apperent deficit of hydrogen and helium. These features suggests that the progenitor stars have been stripped and suffered significant mass-losses before exploding, in addition to revealing large ejecta velocities. Since then, several similar associated events have been observed and the GRB-SN connection is now firmly established \citep{Stanek03,Hjorth03,Woosley06,Cano17}. A typical spectral evolution of a long GRB transitioning into a SN type Ic-BL is shown in Fig.~\ref{fig:grbsngwkn} for the recent GRB\,180728A (Rossi et al., in preparation). There are, however, also examples of long GRBs with apparently no associated SNe \citep[e.g.][]{Fynbo06b,DellaValle06,Gehrels06}. Conversely, several SN Ic-BLs have also been detected without being associated with GRBs \citep{Soderberg10,Modjaz16}. While the jet-structured geometry of the GRB output certainly plays a role in the detection of the prompt $\gamma$-ray emission in these cases, some jets are also expected to be \lq choked\rq~in the interior of the progenitor star before breaking out. In this process, a significant fraction of the energy from the jet is dissipated into a hot \lq cocoon\rq. The first evidence for this was only observed recently for GRB\,171205A \citep{Izzo19}, demonstrating that at least some GRB-less SNe could be explained by a choked jet scenario.

In the scenario where short GRBs are produced by the merger of two neutron stars, 
the expected associated transient is a kilonova \citep[e.g.][]{Metzger12}. The first tentative detection of a kilonova following a short GRB was observed as an excess in the late-time near-infrared light curve of GRB\,130603B \citep{Tanvir13,Berger13}. The smoking gun to the origin of short GRBs came from the kilonova AT\,2017gfo accompanying the gravitational wave (GW)-detected neutron star merger GW\,170817 \citep{Abbott17a}. Only 1.7 seconds after the GW observations of the coalescence of the binary neutron star system, the {\it Fermi} satellite detected a new short GRB \citep{Goldstein17} spatially consistent with the GW event. The optical counterpart was identified less than 11 hours after the merger \citep{Abbott17b}, and the first spectroscopic observations started after roughly 35\,hours \citep{Smartt17,Pian17}. The first four epoch spectra observed with the X-shooter spectrograph mounted on the ESO-VLT are shown in Fig.~\ref{fig:grbsngwkn}, kindly provided by the ENGRAVE collaboration\footnote{\url{www.engrave-eso.org}}. The spectral evolution reveals a rapidly-fading blue transient that quickly becomes significantly reddened, showing line features that are consistent with light $r$-process elements. This benchmark case provided conclusice evidence that binary neutron star mergers produce gravitational waves, in addition to electromagnetic radiation in the form of short GRBs and radioactively powered kilonovae.

\subsection{Quasar classification and selection} \label{sec:qsosel}

Intrinsically, the emitting spectra of quasars follow a similar spectral power-law slope as observed for GRB afterglows and is also believed to be produced by synchrotron radiation. Quasars are typically observed with very strong energy output, ranging from X-ray to radio wavelengths, and characteristic broad (Type I) or narrow (Type II) emission lines. They were first detected as radio sources with a seemingly star-like optical counterpart, hence the name quasi-stellar radio source or quasar \citep{Schmidt63,Matthews63}. Because of their distinct spectral shape, they were also found to appear more \lq blue\rq~than typical main-sequence stars and could therefore be selected in greater numbers without relying on radio detections based on their ultraviolet excess \citep[UVX;][]{Sandage65,Schmidt83}. This approach has been used to build extensive quasar surveys such as the Sloan Digital Sky Survey \citep[SDSS;][]{York00} or the 2dF QSO redshift survey \citep[2QZ;][]{Croom04} and the number of spectroscopically-confirmed quasars now counts more than $500,000$ individual sources \citep[e.g.][]{Paris18}.

However, the UVX selection technique has two major disadvantages. First, for quasars located at redshifts above $z\gtrsim 2.2$, the Lyman-$\alpha$ (\lya) forest starts to enter the optical $u$-band, effectively removing the observed $u$-band excess. Reddening from dust located either in the quasar host or in an intervening system (see also Sect.~\ref{ssec:dust}) will also extinguish most of the emitted light, especially in the bluest part of the spectrum, again diminishing the UV excess. To circumvent these two disadvantages, more efficient quasar selection techniques based on near-infrared \citep{Warren00} and mid-infrared \citep{Stern12} photometry were introduced. 

In general, to gauge the true underlying population of quasars it is important to study a well-defined sample, free of selection bias. One approach is to target the specific sub-population of quasars that are missed in most current optical surveys, by defining a set of tailored selection criteria based on optical to near/mid-infrared photometry \citep[see e.g.][]{Fynbo13a,Krogager15,Krogager16b}. This type of selection forms the basis of Chapter~\ref{chap:kvrq}. Another approach is to redefine the selection criteria and build a complete and unbiased sample of quasars that is not based on any intrinsic properties. Chapter~\ref{chap:gaiasel} presents such a novel selection, where we identify quasars solely as sources with zero proper motion on the sky using the astrometric measurements from the on-going {\it Gaia} mission.

\subsection{GRBs and quasars as cosmic probes}

Due to their immense intrinsic brightness, GRBs and quasars offer a viable approach to study the faintest galaxy population and the high-redshift Universe. The most distant observable structures (except for the CMB) are the population of galaxies and quasars at $z \sim 7 - 10$, at which point the reionization epoch and the formation of the first stars and galaxies initiated. To date, only two quasars have been identified at $z > 7$ \citep{Mortlock11,Banados18}. The most distant GRBs found have been GRBs\,090423 and 120923A, with spectroscopic redshifts of $z = 8.2$ \citep{Salvaterra09,Tanvir09} and $z=7.8$ \citep{Tanvir18}, respectively. Two GRBs have also been found to have photometric redshifts of approximately $z=9.4$ \citep{Cucchiara11} and $z=7.9$ \citep{Bolmer18}, although with larger uncertainties associated with placing the exact location of the {\lya} break. While the number of GRBs and quasars at the epoch of first star formation is still scarce, the fact that they can in principle be observed at this time makes them valuable tools to study galaxies through most of cosmic time.

\begin{figure}[!t]
	\centering
	\includegraphics[width=12cm]{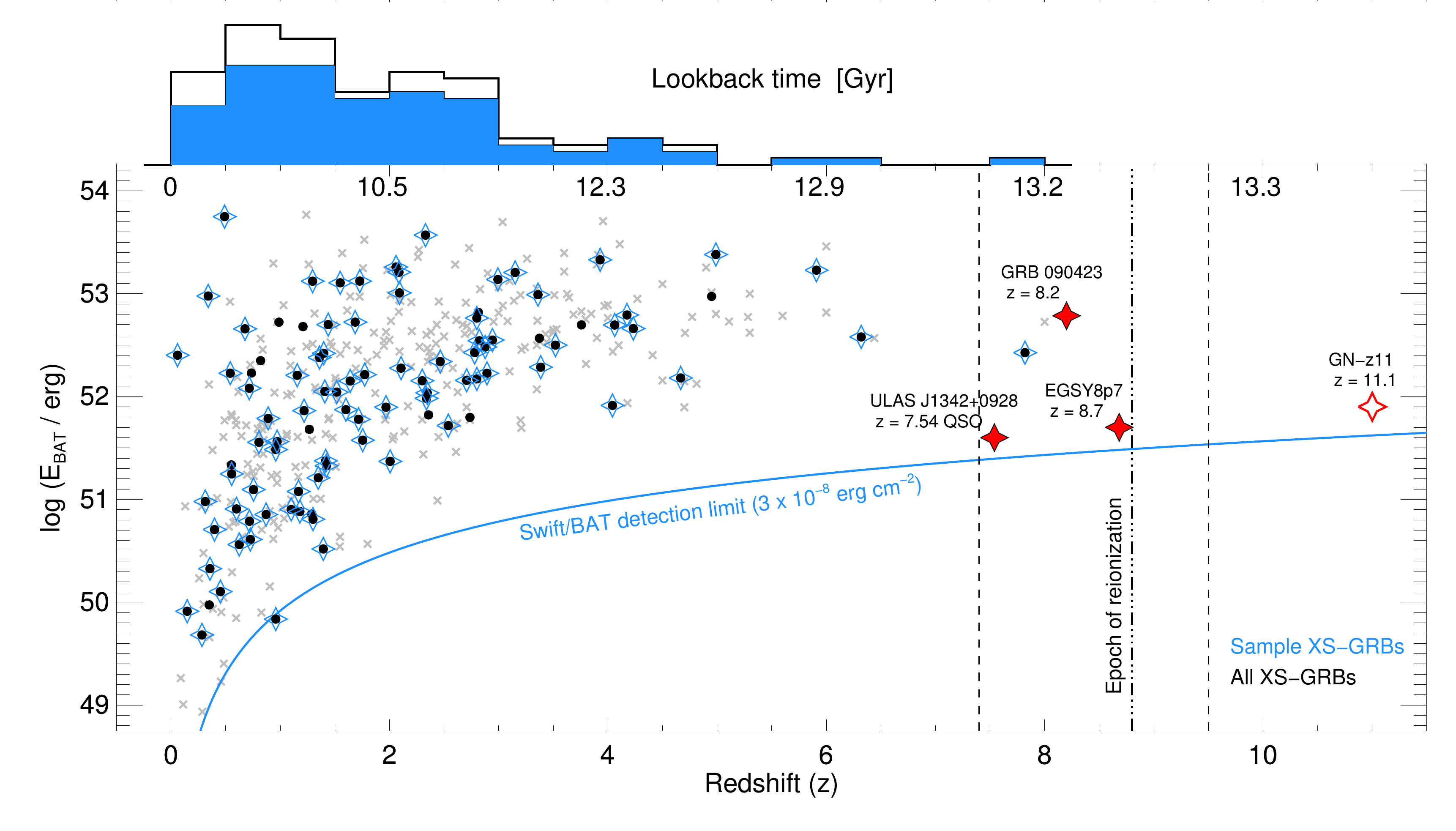}
	\caption{Redshift distribution of all {\it Swift} GRBs with measured redshifts (grey) as a function of intrinsic BAT $\gamma$-ray energy, $E_{\rm BAT}$. Bursts that are part of the statistical XS-GRB sample are marked by the blue stars, whereas black dots represent all GRBs observed with X-shooter. GRBs have been detected spectroscopically up to $z\approx 8$. Overplotted as the red stars are the GRB \citep{Tanvir09,Salvaterra09}, quasar \citep{Banados18}, and galaxies \citep{Zitrin15,Oesch16} with the highest spectroscopically confirmed redshifts, the latter three shown at arbitrary $E_{\rm BAT}$. The blue solid line represents approximately the BAT sensitivity limit and it is clear that only a fraction of the most energetic GRBs are detected at high redshift. The estimated epoch of reionization is shown by the black dot-dashed line, with the uncertainty shown as the black dashed lines. Both GRBs and quasars have the potential to probe the first stars and galaxies formed, even as early as during the reionization epoch.
	This figure was produced for and first presented by \citet{Selsing19}.}
	\label{fig:grbz} 
\end{figure}

Due to their transient nature, it is important to have efficient selection criteria for GRBs to optimize the follow-up effort. Extensive samples of quasars are in that regard more easily constructed, whereas GRBs offer the advantage of being much less susceptible to dust obscuration. To maximize the scientific yield from GRB samples it is therefore important to define an unbiased selection of bursts that are still broad enough to include the majority of the underlying GRB population. Most GRB afterglow surveys have therefore been defined in such a way that only exclude bursts based on conditions local to the Milky Way but independent of intrinsic GRB properties \citep[e.g.][]{Jakobsson06a,Fynbo09}. Building on this, the X-shooter GRB (XS-GRB) afterglow legacy survey was defined \citep{Selsing19}, that now counts more than 100 optical/near-infrared GRB afterglows, which a large fraction of the work done in this thesis is based on. The redshift distribution as a function of energy, $E_{\rm BAT}$ \citep[defined as $E_{\rm BAT} = F_\gamma\,4\pi\,d_L^2(1+z)^{-1}$;][]{Lien16}, of the full sample of XS-GRBs are shown in Fig.~\ref{fig:grbz}. It is clear that at large redshifts, only a fraction of the brighest GRBs can be detected by {\it Swift} but that the XS-GRB follow-up campaign targets the underlying {\it Swift}-detected GRB population homogeneously. Before drawing any conclusions related to the characteristics of the underlying population of GRB hosts it is, however, important to understand how the physical properties such as star-formation rate (SFR), stellar mass and metallicity might also influence the GRB production rate. It was initially believed that GRB hosts should follow the general star formation history \citep{Wijers98}. However, specifically at low redshifts ($z < 1.5$), GRBs have been shown to occur preferentially in low-metallicity environments \citep{Kruhler15,Schulze15,Japelj16,Vergani17}, translating into generally lower stellar masses and fainter luminosities for their host galaxies \citep{Sollerman05,Wolf07,Vergani15,Perley13,Perley16b}. It is now evident, however, that luminous, massive and hence metal-rich GRB host galaxies do exist but they are often associated with dusty or \lq dark\rq~\citep[e.g.][]{Jakobsson04} GRB afterglows \citep{Kruhler11,Rossi12,Perley13,Hunt14} and are therefore as a consequence underrepresented in samples selected by optical afterglow identification.

Material along the line of sight to both GRBs and quasars might also influence the detection probability. Even the faintest intervening or host galaxies are for example discovered by their associated absorption signatures imprinted on the background emission spectrum. These absorption systems are the main focus of this thesis, and a more detailed introduction to the phenomena are given below.

\section{Damped Lyman-$\alpha$ absorbers} \label{sec:dlas}

Previously, the study of the high-redshift galaxy population was limited to the study of intervening absorption systems observed toward background quasars \citep{Weymann81}. While the capabilities of direct imaging of high-$z$ galaxies has recently undergone a revolution \citep{Madau14,Stark16}, the use of first quasars \citep[and later GRBs;][]{Kulkarni98} as cosmic tools to illuminate even the faintest high-$z$ foreground or host galaxy absorption systems still provide the most detailed information about this galaxy population \citep{Wolfe05}. This approach is also extremely powerful, since it can easily be extended out to the epoch of reionization (as mentioned above) and is not biased towards the brightest or most massive galaxies as is the case for the typical UV-selected galaxy surveys.

Detailed absorption-line analyses in GRB host galaxies have been possible for systems as far as $z\approx 6$ \citep{Kawai06,Hartoog15}. Currently, nine GRB host absorption systems within the first Gyr after the Big Bang ($z \gtrsim 4.7$) have measured gas-phase abundance, where in the same redshift range, slightly fewer than 20 quasar absorption systems have been identified \citep{Rafelski14,Poudel18}. Identifying galaxies in absorption is typically based on the strong {\lya} line transition from neutral hydrogen (\hi) in the line of sight, which implies a large amount of neutral gas. However, in sightlines towards very high-$z$ quasars ($z \gtrsim 5.7$), this approach becomes increasingly more difficult due to the large increase in the Gunn-Peterson optical depth \citep{Fan06}. Combined with significant blending with the {\lya} forest whose line density per unit redshift increases with redshift, the flux bluewards of the quasar {\lya} emission line will be completely suppressed, thus concealing any traces of neutral hydrogen absorbers \citep[though identifying the associated metal line transitions at longer rest-frame wavelengths is still possible;][]{Cooper19}. Identifying the host galaxy counterparts of GRB absorption systems is not affected by this effect though, since the {\lya} line is at the same redshift as the afterglow emission component so the {\lya} forest will only appear bluewards of {\lya} at the redshift of the GRB host galaxy. On the other end, the {\lya} line transition at $\lambda_{\rm rest} = 1215.67$\,{\AA} is only observable from the ground when located at redshifts above $z\gtrsim 1.7$ due to the atmospheric cut-off. The detailed information that can be obtained from absorption-selected galaxies are therefore mainly limited to galaxies in the redshift range $z \sim 2 - 6$.

\begin{figure}[!t]
	\centering
	\includegraphics[width=12cm]{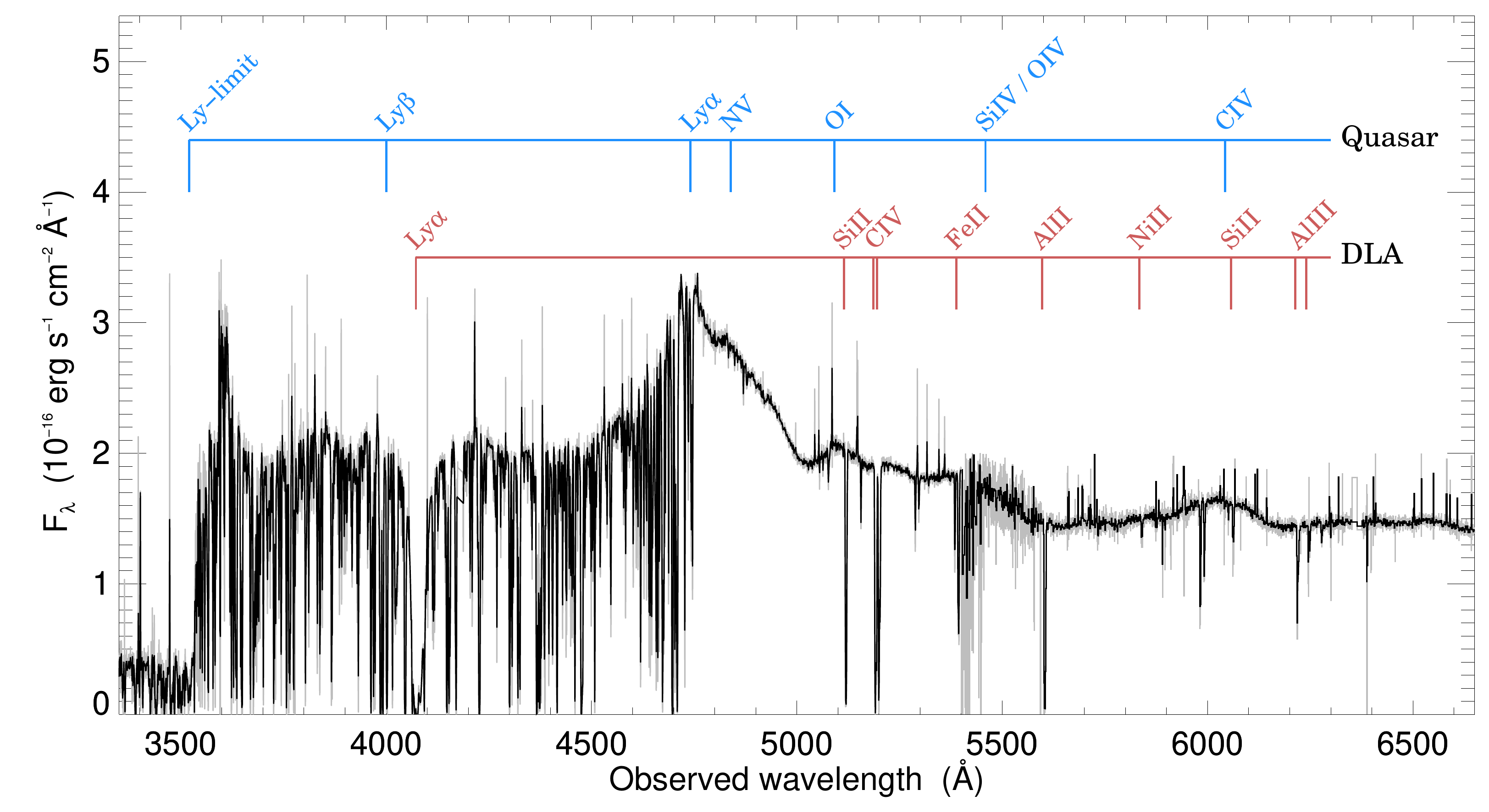}
	\caption{Spectrum of the quasar Q\,2222-0946 at $z_{\rm em} = 2.93$ with an intervening DLA at $z_{\rm DLA} = 2.35$ \citep[from][]{Fynbo10}. The most prominent emission lines from the quasar and absorption lines from the DLA are marked in blue and red, respectively. The strong Ly$\alpha$ feature is clearly identified in the Ly$\alpha$ forest.}
	\label{fig:qsodla} 
\end{figure}

The strongest of the neutral hydrogen absorbers are called damped {\lya} absorbers (DLAs) and are defined by having $N$(\hi) $> 2\times 10^{20}$\,cm$^{-2}$. An example of an intervening DLA in a typical quasar spectrum is shown in Fig.~\ref{fig:qsodla}. The broad {\lya} absorption trough can be identified even in low-resolution spectra and are easily distinguished from the typical narrow {\lya} forest lines. With the goal of identifying the neutral-gas disks of high-redshift galaxies, \citet{Wolfe86} carried out the first survey for this particular type of absorbers in quasar spectra. These systems dominate the cosmic reservoir of neutral gas from $z\approx 0 - 5$ \citep{Noterdaeme09a,SanchezRamirez16} and contain sufficient mass at $z>3$ to account for most of the visible stellar mass in present day galaxies \citep{Storrie00}. While the defined {\hi} column density threshold for DLAs appear arbitrary, it ensures that a large fraction of the gas is neutral and thus completely self-shielding \citep[see e.g.][]{Wolfe05}. This means that ionization corrections are negligible, such that the singly-ionized metal lines with ionisation potentials below 13.6 eV trace the total metal abundance [i.e. $N$(Zn\,{\sc ii}) / $N$(\hi) = $N$(Zn) / $N$(H)]. The high neutrality of the gas is a vital precursor for the formation of molecular clouds and subsequent star formation. Other {\lya} absorbers include sub-DLAs ($10^{19} < N$(\hi) $< 2\times 10^{20}$\,cm$^{-2}$), Lyman-limit systems (LLS; $10^{17} < N$(\hi) $< 10^{19}$\,cm$^{-2}$), and Ly$\alpha$ forest absorbers ($N$(\hi) $< 10^{17}$\,cm$^{-2}$), which all contain hydrogen mainly in ionized form \citep{Rauch98,Prochaska99,Wolfe05}. DLAs therefore provide one of the most accurate probes of the chemical enrichment of the neutral gas-phase of the ISM in galaxies through most of cosmic time.

Since their first discovery, multiple surveys for DLAs in quasar sightlines have been carried out \citep{Lanzetta91,Lanzetta95,Wolfe95,Boisse98,Storrie00,Rao00,Ellison01,Peroux03,Rao06}. With the contribution of the extensive spectroscopic database from the SDSS, several thousand DLAs have been identified in quasar sightlines to date \citep{Prochaska04,Prochaska05,Noterdaeme09a,Noterdaeme12b}, with a few hundred observed with high-resolution spectrographs \citep{Ledoux03,Ledoux06,Noterdaeme08,Rafelski12,Berg15,DeCia16}. 

\begin{figure}[!t]
	\centering
	\includegraphics[width=12cm]{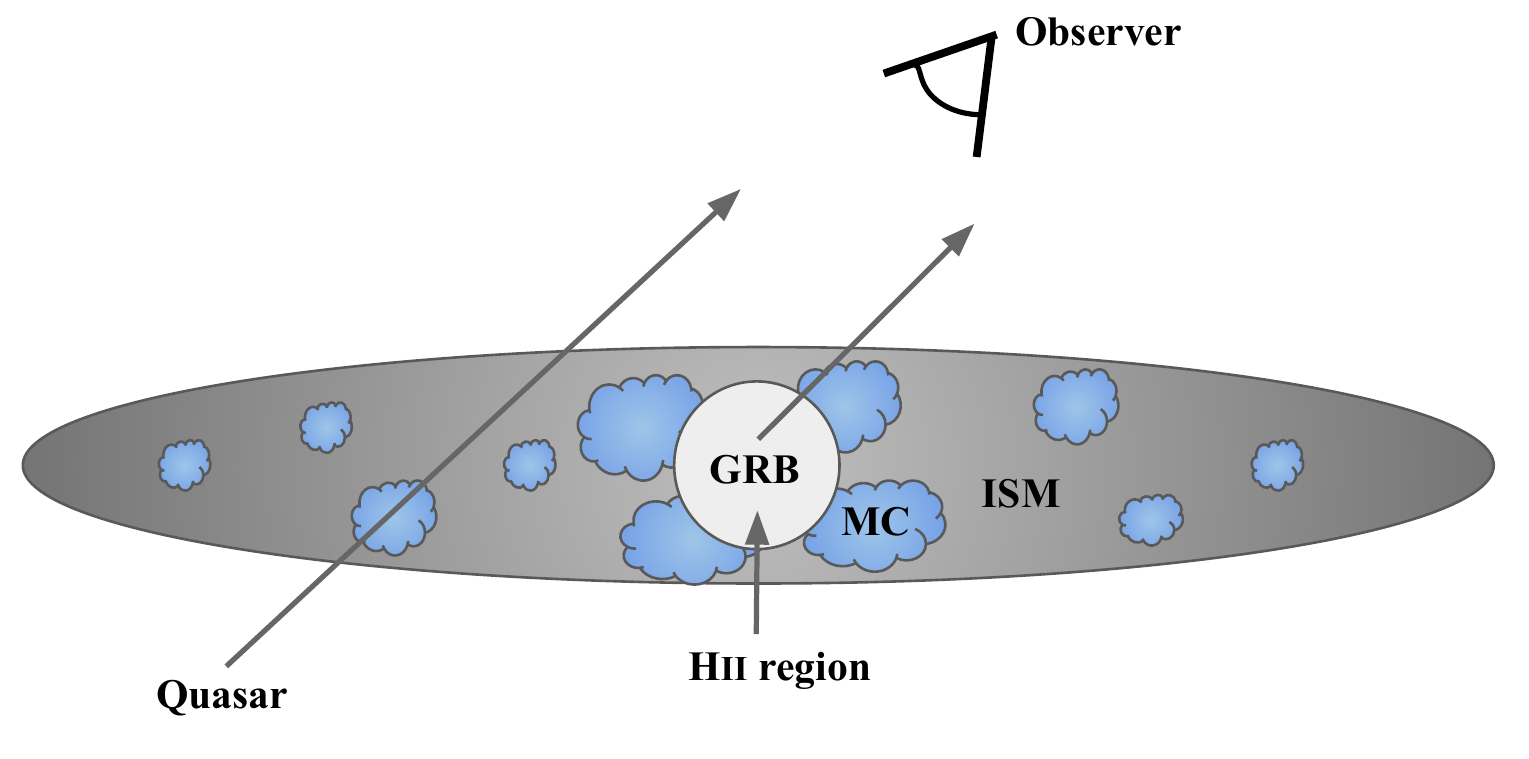}
	\caption{A simplistic cartoon illustration of the typical impact parameters for GRB and quasar sightlines \citep[similar to fig. 1 from][]{Prochaska07}. The extent of the galactic disc is roughly on a scale of 10\,kpc. GRB absorption systems are expected to originate close to the center of the host, whereas quasar sightlines are more likely to probe the outer regions of the ISM. The typically observed larger abundance of {\hi} in GRB absorbers is purely a selection effect. The blue bubbles represent cold molecular clouds distributed throughout the host galaxy environment. The \lq clumpiness\rq~increases closer to the center such that the detection probability of H$_2$ and {\ci} decreases as a function of impact parameter.}
	\label{fig:grbqsodla} 
\end{figure}

Significant effort has also been put into follow-up campaigns of GRBs, in which DLAs are typically detected in the optical afterglow spectra \citep{Fynbo09,Selsing19}. More than 100 neutral hydrogen absorbers have now been identified in GRB host absorption systems, with the majority ($\gtrsim 85\%$) being classified as DLAs \citep{Tanvir19}. The main difference between the host galaxy absorbers observed in GRB afterglows compared to the absorption systems intervening quasar sightlines is that the former population is observed to be substantially more gas-rich on average \citep{Vreeswijk04,Jakobsson06b,Watson06}. Comparing the two absorber populations requires a good understanding of the underlying selection bias introduced when building the respective samples. While GRBs are likely to explode within the central regions of their host galaxies, quasar absorbers are selected depending on their neutral gas cross-section which is related to their size and luminosity \citep{Zwaan05}. As a consequence, GRB host galaxy absorption systems are expected to show higher gas and metal column densities than quasar absorbers, as is observed \citep{Savaglio03,Savaglio06,Prochaska07,Fynbo08,Fynbo09,Arabsalmani15}. A schematic of this sightline effect is shown in Fig.~\ref{fig:grbqsodla}. From the figure it is clear that quasar absorbers tend to probe the outer regions of the ISM or the halo gas of the intervening galaxy, whereas GRB sightlines probe the hearts of their star-forming host galaxies.

\subsection{Gas-phase abundances and kinematics}

In addition to the strong {\lya} absorption line typically observed in GRB and sporadically in quasar spectra, associated absorption lines from metal transitions are also commonly detected. By deriving the column densities of the observed metal line transitions, it is possible to infer the gas-phase abundances of a specific element, X, in the line of sight. If the column density of neutral hydrogen can additionally be constrained, the measured relative abundances can be compared with the equivalent solar values, providing the metallicity of the absorption system as 
\begin{equation*}
{\rm [X/H]} \equiv \frac{\log N({\rm X})}{\log N({\rm H})} - \frac{\log N({\rm X})_{\odot}}{\log N({\rm H})_{\odot}}
\end{equation*}
where [X/H] = 0 defines solar metallicity of X \citep[with reference values most widely adopted from][]{Asplund09}. Again, since the gas is predominantly neutral in DLAs, the low-ionization transitions will trace most of the element X abundance. To obtain a reliable estimate of the actual gas-phase metallicity, only volatile or mildly depleted elements such as zinc or sulphur should be used as tracers. Some fraction of the metals in the ISM are also expected to be in the dust-phase, such that they will appear depleted from the observed abundances. By estimating the depletion pattern it is possible to correct for this effect, however, and accurately derive the total metal content of DLAs (see Sect.~\ref{ssec:dust}), in addition to estimate the dust content of the absorbing system.

\begin{figure}[!t]
	\centering
	\includegraphics[width=12cm]{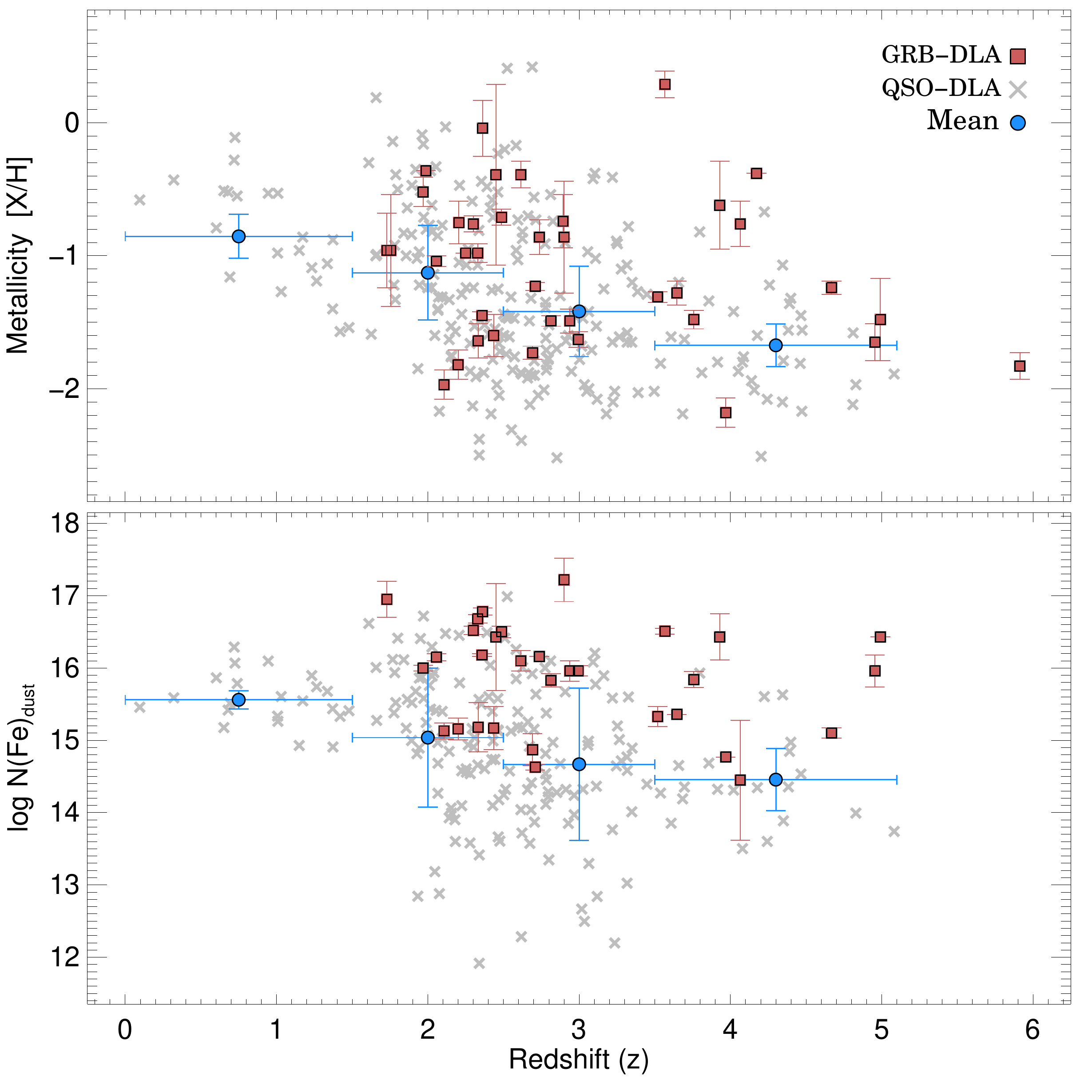}
	\caption{Redshift evolution of neutral gas-phase metallicity (upper panel) and dust-phase iron column density (lower panel). Grey crosses represent a recent compilation of quasar DLAs \citep{Berg15,DeCia18} and red squares denote GRB absorbers \citep[e.g.][]{Bolmer19}. The blue dots represent the average quasar DLA metallicity divided into redshift bins of $0 < z < 1.5$, $1.5 < z < 2.5$, $2.5 < z < 3.5$, and $z > 3.5$ (with the sample variance as the vertical error bars). Both absorber populations are found to show an increase in metal and dust content for decreasing redshifts.}
	\label{fig:metz} 
\end{figure}

Since DLAs are believed to be the dominant reservoirs of neutral gas at all redshifts, they are unique tools to map out the cosmic chemical enrichment of the ISM in galaxies through most of cosmic time. For instance, they have been widely used to trace the redshift evolution of the gas-phase metallicity in quasar absorbers \citep{Pettini94,Pei99,Kulkarni02,Prochaska03,Rafelski12,Rafelski14,DeCia18} and GRB host absorption systems \citep{Fynbo06a,Cucchiara15}, which both agree on a common evolution of the average gas-phase metallicity as a function of redshift.
The latest compilation of quasar DLAs by \cite{Berg15}, observed with high-resolution spectrographs (following \citealt{DeCia18}), for which reliable gas-phase metallicies could be derived is shown in the top panel of Fig.~\ref{fig:metz}. Overplotted are the GRB host absorption systems observed with medium- to high-resolution spectrographs \citep[the largest contribution being from the XS-GRB sample;][]{Selsing19} for which an accurate metallicity could be derived. Both absorber populations appear to follow similar trends of increasing metallicities with decreasing redshift.

One important caveat to keep in mind before drawing conclusions for the underlying absorber population is the effect of selection bias, and in particular if a certain type of absorption system is excluded from existing samples. While the high-energy emission from GRBs is virtually unaffected by dust, very dust-obscured sightlines will significantly hamper the spectroscopic follow-up of their optical afterglows and will thus appear as optically \lq dark\rq~bursts \citep{Fynbo01,Fynbo09,Jakobsson04}. Typical quasar selection techniques are also subject to a significant dust bias, excluding sightlines with intervening dusty and therefore likely also metal-rich foreground galaxies \citep{Fall93,Pontzen09,Krogager19}. We attempt to quantify this selection bias in Chapter\,\ref{chap:dustydla}, for both the quasar and GRB absorber populations.

If the early assumptions that the broad {\lya} absorption feature observed toward quasars trace the gaseous, thin disk of proto-type galaxies \citep{Wolfe86} is true, then this should also be reflected in the velocity structure seen in the metal line profiles. One approach to quantify the velocity width of the low-ionization lines is by measuring $\Delta V_{90}$, defined as the velocity encompassing 90\% of the apparent optical depth, $\tau$ \citep{Prochaska97,Ledoux98}. The velocity width can then be calculated as 
\begin{equation*}
\Delta V_{90} = c \frac{\lambda_{95}-\lambda_{5}}{\lambda_{0}}
\end{equation*}
where $\lambda_{0}$ is the line centroid and $\lambda_{5}$ and $\lambda_{95}$ are the 5$^{\rm th}$ and 95$^{\rm th}$ percentiles of the distribution of $\tau$ for a given line. Based on models including $\Delta V_{90}$ and the velocity structure of the absorption line profiles, \citet{Prochaska97,Prochaska98} concluded that the metal lines observed in quasar DLAs are consistent with probing large rotating disks. In addition, \citet{Ledoux06} also found a relation between $\Delta V_{90}$ and [X/H], which they argued was a consequence of an underlying mass-metallicity relation for the galaxies responsible for the DLA absorption lines. This suggests that the quantity $\Delta V_{90}$ holds information about the dynamical mass of the absorbing galaxies.

The dominant metal lines from DLAs observed in quasar and GRB afterglow spectra are from low-ionization line transitions, used to constrain the metallicity and kinematics of the neutral gas as described above. Most absorbers also show lines from higher ionization states such as C\,{\sc iv}, Si\,{\sc iv}, N\,{\sc v} and O\,{\sc vi} which is believed to trace the hot circumgalactic halos \citep[e.g.][]{Fox07a}. The high-ionization lines are typically observed to show larger velocity widths than the low-ionization lines, indicating a larger velocity dispersion in the hot photoionized gas component compared to the neutral gas. Some of these high-ionization lines might also be populated by enhanced ionization if an absorber is close to the background quasar \citep[typically designated as \lq proximate\rq~DLAs,][]{Ellison10}
or from the increased photon flux from the prompt and afterglow emission of the GRB \citep{Prochaska08b,Fox08}. There is at least now conclusive evidence that the GRB can have an effect on the surrounding environment. Based on the detection and line variability of excited fine-structure lines, \citet{Vreeswijk07} determined that the excited lines are populated by UV-pumping from the GRB afterglow. Based on this model, it is also possible to infer the distance from the GRB to the absorbing cloud (which is typically found to be of the order $d = 0.5 - 2$\,kpc) based on the radiation field required to populate the excitated transitions. 

\begin{figure}[!t]
	\centering
	\includegraphics[width=12cm]{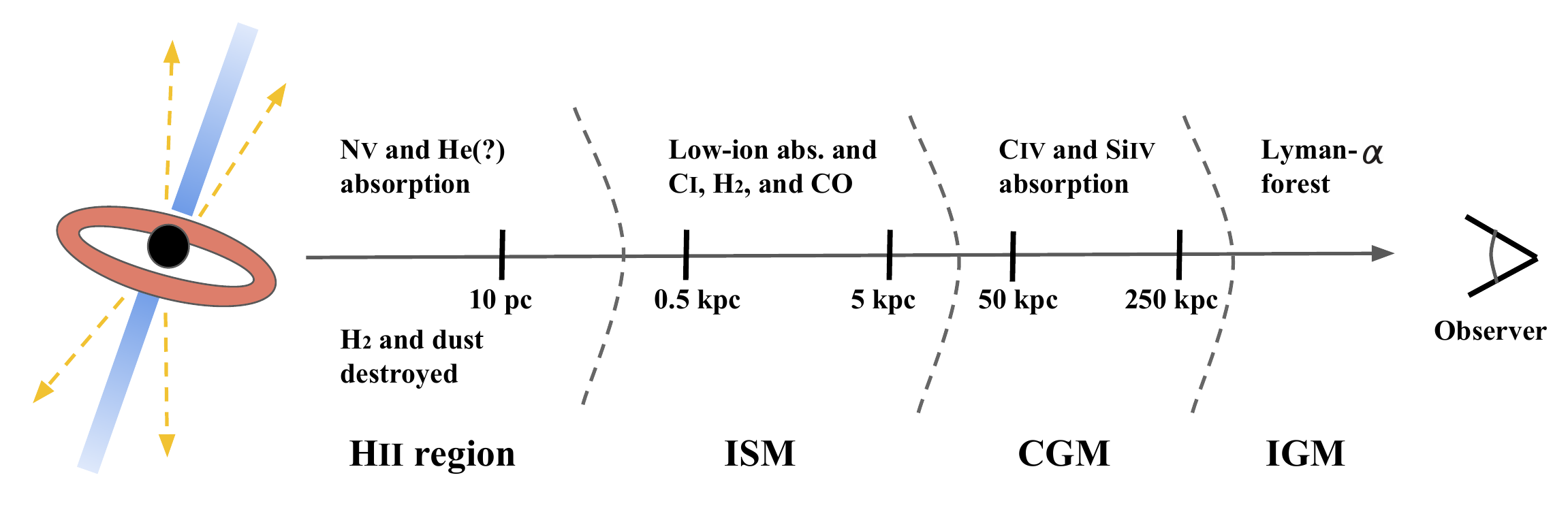}
	\caption{Schematic showing the location of the different absorption components in the line of sight to a GRB (loosely based on fig. 2.9 from \citealp{ThoenePhD}). Close to the burst ($d \lesssim 10$\,pc), dust and H$_2$ molecules will be destroyed. N\,{\sc v} is expected to be produced here (either from photo-ionization or recombination) and significant He absorption of soft X-ray photons are expected here as well. At $d \approx 0.5 - 2$\,kpc the main absorption components from the low-ionization metal lines and fine-structure transitions probe the neutral ISM. Further out at $d \approx 50 - 250$\,kpc the high-ionization lines C\,{\sc iv} and Si\,{\sc iv} are though to probe the hot halo of the cirgumgalactic material. At high-$z$, absorption components from the intergalactic medium will be present as well.}
	\label{fig:losprop} 
\end{figure}

In Fig.~\ref{fig:losprop} is shown a schematic of the different locations of the both the low- and highly-ionized absorbing gas, as observed in the sightline to the GRB from the afterglow spectra. While the majority of the absorbing gas might be located at several kpc from the GRB, a circumburst origin for some of the observed features might not be completely ruled out. For example, \cite{Prochaska08b} argue that N\,{\sc v} could be populated by photoionization from the afterglow to within $d\lesssim 10$\,pc from the burst. It is also on the same scale that helium is expected to cause most of the observed soft X-ray absorption \citep{Watson13}. Connecting these two observables is the focus of Chapter\,\ref{chap:highion}, where a systematic study of the circumburst gas and metals surrounding GRBs is presented as well.

\subsection{The effect of dust} \label{ssec:dust}

\subsubsection*{Depletion of refractory elements}

As described above, the gas-phase abundances of the neutral ISM can be derived by measuring the column density of the element, X, and compare it to the neutral hydrogen content. However, refractory elements such as iron typically show very low relative abundances compared to the mildly depleted elements such as zinc and sulphur. For any refractory element, Y, the depletion level can be derived relative to a volatile element, X, as
\begin{equation*}
{\rm [X/Y]} = \frac{\log N({\rm X})}{\log N({\rm Y})} - \frac{\log N({\rm X})_{\odot}}{\log N({\rm Y})_{\odot}}
\end{equation*}
The depleted fraction of a given metal in the dust phase is then $D({\rm Y}) = 1-10^{\rm [Y/X]}$, which also represents the dust-to-metals ratio in the absorbing system. The reason why large fractions of refractory elements are missing from the gas-phase is because they are locked and condensed into dust and therefore appear depleted \citep[e.g.][]{Jenkins87,Savage96}. These dust grains are primarily made up of O, Si, Mg, and Fe metals \citep{Draine03}. The strength of the dust depletion depends primarily on the density and chemical enrichment of the gas. For example, \cite{Jenkins09} showed that a wide range of abundances in more than 200 Galactic sightlines are tighly connected to each other via a single parameter, the depletion strength $F_*$. Based on this, \cite{DeCia16} derived empirical sequences for the relative abundances and demonstrated that they are universally valid for sightlines through the Galactic ISM and high-redshift DLA systems.

From the depletion level, it is possible to derive the amount of refractory elements in the dust-phase of the ISM \citep{Vladilo06}. For example, the column density of iron in the dust-phase is given as
\begin{equation*}
N({\rm Fe})_{\rm dust} = N({\rm X})\,\left(1-10^{\rm [Fe/X]}\right)\,\left(\frac{\rm Fe}{\rm X}\right)_{\odot}
\end{equation*}
where X is again any volatile of mildly depleted element. Studying the redshift evolution of $N({\rm Fe})_{\rm dust}$ might provide clues for the production and cosmic evolution of dust. In the bottom panel of Fig.~\ref{fig:metz}, the dust-phase iron column density is shown as a function of redshift for the same samples of quasar and GRB absorbers described in the previous section. It is clear that the GRB absorbers generally show elevated abundances of elements in the dust phase compared to quasar absorbers. $N({\rm Fe})_{\rm dust}$ also appears to follow a similar evolution trend as the metallicity for the same absorption systems. It is important to caution here though that, particularly at high-$z$, the large values of $N({\rm Fe})_{\rm dust}$ especially observed in the GRB absorbers, could also be explained by a deficiency of iron-peak elements which have not had enough time to be produced from type Ia supernovae.

\subsubsection*{Line-of-sight extinction}

\begin{figure}[!t]
	\centering
	\includegraphics[width=12cm]{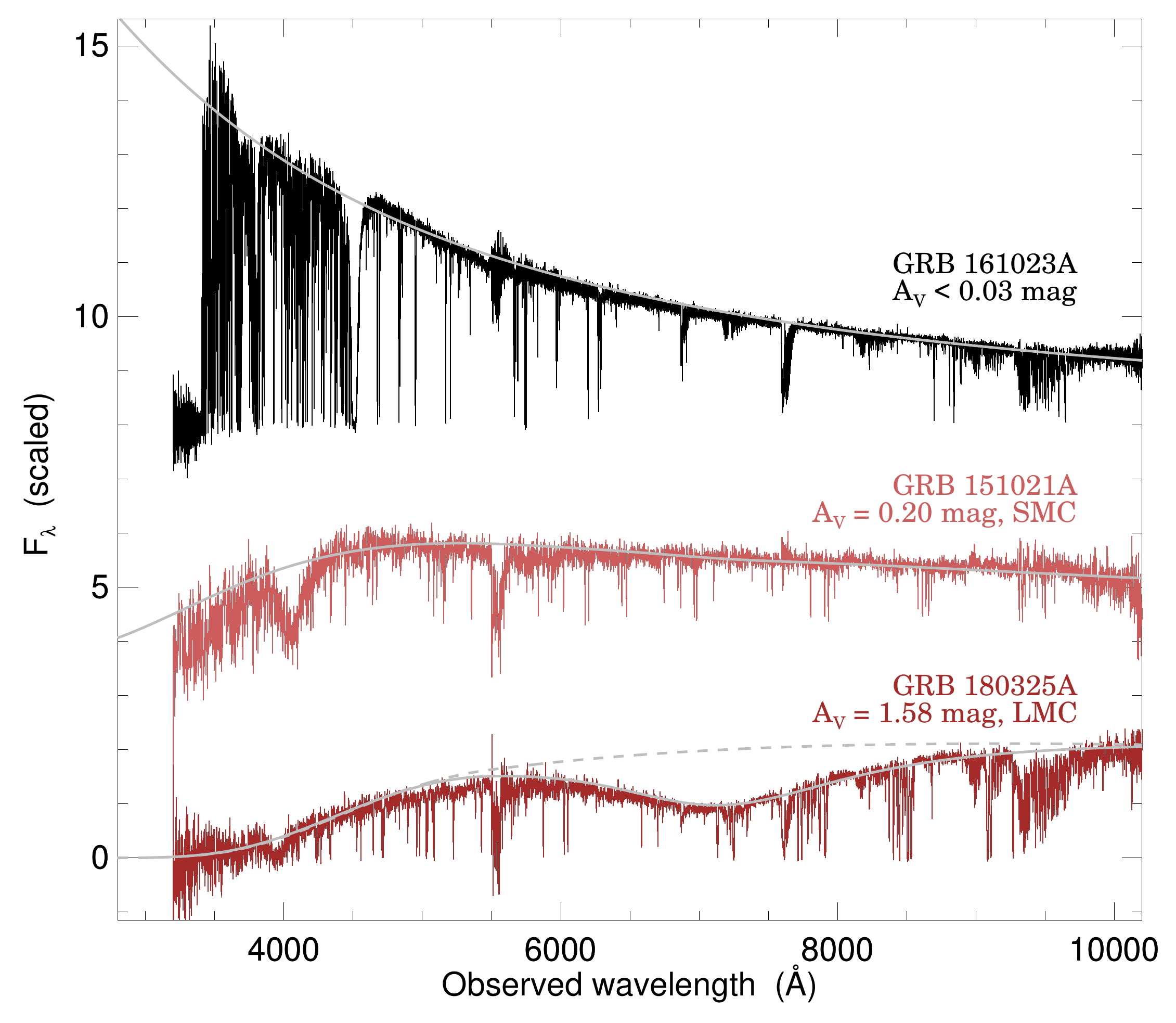}
	\caption{Three distinct optical afterglow spectra of GRBs at $z\approx 2.5$ illustrating the effect of dust. The spectrum at the top is of GRB\,161023A, showing little to no dust extinction \citep[with $A_V < 0.03$\,mag;][]{deUgartePostigo18}. The two bottom spectra show the dust-extinguished optical afterglows of GRB\,151021A (with $A_V = 0.20$\,mag, see Chapter~\ref{chap:cisurvey}) and GRB\,180325A \citep[with $A_V = 1.58$\,mag;][]{Zafar18a}. The latter GRB afterglow also has an additional extinction \lq bump\rq~imprinted on the spectrum, known as the 2175\,\AA~dust extinction feature, which is illustrated by the grey dashed line showing the same extinction curve but without the bump. }
	\label{fig:grbdust} 
\end{figure}

Another observable effect of dust is the scattering of light in the line of sight from the emission source to the observer. The amount of scattering is wavelength-dependent, causing the background source to appear \lq reddened\rq, and also causes an overall dimming effect. The dust {\it extinction} properties of the ISM in a galaxy is distinct from the dust {\it attenuation} properties, which is subject to complex radiative transfer effects reprocessing the stellar light and which also depends on the geometric distribution of the dust, gas, and stars \citep[e.g.][]{Narayanan18}. The shape of the dust extinction curve is only dependent on the dust composition and grain size distribution of the dust particles located in the line of sight. In Fig.~\ref{fig:grbdust}, three different GRB optical afterglow spectra are shown; one showing little to no dust (GRB\,161023A), one with a small amount of dust with properties described by a smooth, SMC-like extinction curve (GRB\,151021A), and one significantly dust-reddened sightline showing an additional extinction \lq bump\rq~(GRB\,180325A), known as the 2175\,\AA~dust extinction feature \citep[e.g.][]{Draine89}. Carbonaceous dust grains are believed to produce this particular type of extinction \citep{Henning98}, and the overall shape of the extinction curves can be described by a combination of silicates and carbonaceous material \citep{Draine03}.

The extinction (or reddenening) curve is typically quantified by the wavelength-dependent extinction $A_{\lambda}$, relative to the $V$-band extinction, $A_V$. The extinction can also be expressed in terms of the color excess from the $B$- to $V$-band extinction, $A_B - A_V = E(B-V)$, which is linked to the $V$-band extinction via the total-to-selective reddening paramater, $R_V = A_V / E(B-V)$. This quantity is then simply a measure of the slope of the extinction curve, which is related to the average dust grain size distribution (where larger particles result in higher observed values for $R_V$). Previously, the most commonly adopted extinction curve models was parametrized by \cite{Cardelli89} and \cite{Pei92}. A more general formulation was provided by \citet{Fitzpatrick90,Fitzpatrick07}, which will be used throughout this thesis, that describes the extinction curve through a set of nine parameters as
\begin{equation*}
A_{\lambda} = \frac{A_V}{R_V}\,\left(k(\lambda-V) + 1\right)~,
\end{equation*}
where the relative reddening, $k(\lambda-V)$, is given as
\begin{equation*}
k(\lambda-V) = \begin{cases} 
c_1 + c_2x + c_3D(x,x_0,\gamma), &\mathrm{for}~x\le 5.9  \\
c_1 + c_2x + c_3D(x,x_0,\gamma) + c_4(x-c_5)^2, &\mathrm{for}~x > 5.9
\end{cases}~,
\end{equation*}
and the Lorentzian-like \lq Drude\rq~profile representing the 2175\,\AA~dust extinction feature is described as
\begin{equation*}
D(x,x_0,\gamma) = \frac{x^2}{(x^2-x^2_0)^2 + x^2\gamma^2}
\end{equation*}
with $x=\lambda^{-1}$ in units of $\mu$m$^{-1}$. Basically, this dust model contains two components, one describing the linear UV part of the spectrum via the components $c_1$ (intercept), $c_2$ (slope) and the terms $c_4$ and $c_5$ describing the far-UV curvature. The second component is the Drude profile representing the 2175\,\AA~extinction bump, controlled by the parameters $c_3$ (bump strength), $x_0$ (central wavelength) and $\gamma$ (width of the bump). The last two parameters are the visual extinction, $A_V$, and the total-to-selective reddening, $R_V$. The advantage of such a parametrization of the dust-extinction model, is that it for example allows for the strength and width of the 2175\,\AA~dust bump to be constrained independently, and also provides a way for dust extinction curves with no local analogs to be constructed, which we applied in Chapter~\ref{chap:140506a}. The observed dust-reddened spectrum can then be described by a simple model as $F_{\rm obs} = F_{\lambda}\times 10^{-0.4A_{\lambda}}$. The visual extinction, $A_V$, can then be estimated assuming a given dust-extinction model when knowing the redshift of the GRB or quasar absorption system and by assumming some underlying intrinsic spectrum. In the case of GRB afterglows, a smooth power-law of the form $F_{\lambda}=\lambda^{-\beta}$ is typically observed, and is believed to originate from synchrotron radiation emitted from the interaction and deceleration of the ultra-relativistic GRB jet in an external medium \citep[][see also Sect.~\ref{ssec:grbaft}]{Sari98}. To substitute as the intrinsic quasar spectrum, a composite of bright, instrinsically \lq blue\rq~quasar spectra are usually applied, which will be demonstrated in Chapter~\ref{chap:dustydla}. 

If the same dust composition and distribution of the overall ISM is probed by either directly measuring the line-of-sight extinction or as inferred from the relative depletion patterns, the dust content is expected to scale. There are, however, an observed discrepancy between the two. One explanation could be that GRB sightlines probe grey dust and the SED-derived $A_V$ will therefore be understimated \citep{Savaglio03}. Another could be that the two approaches probe dust on different scales and by distinct elements. Reconciling this discrepancy forms the basis of Chapter~\ref{chap:cidust}, where the latter scenario is demonstrated to, at least partially, cause some of the tension.

\subsection{Cold and molecular gas} \label{ssec:cmgas}

Since the neutral gas in DLAs provide a natural \lq fuel\rq~for star formation, molecular hydrogen is expected to be present in most sightlines. The detections of H$_2$ in quasar DLA spectra are, however, found to be scarce with an estimated detection rate of $\lesssim 10\%$ for the overall population of DLAs \citep[see e.g.][]{Jorgenson14,Balashev14,Balashev18}. The detection probability is found to increase for sightlines rich in H\,{\sc i} gas though \citep{Noterdaeme15b}, or showing high metallicities \citep{Petitjean06}, large dust-to-gas ratios \citep{Ledoux03} and significant dust-phase iron column densities \citep{Noterdaeme08}. 

Since GRBs are expected to probe star-forming galaxies, and star-formation rates are driven by the availability of dense, molecular gas, the detection of H$_2$ absorption features was also expected to be a common occurence in the optical afterglow spectra of GRBs. This was also motivated by their large H\,{\sc i} column densities probing the same regimes as observed for H$_2$-bearing quasar DLAs \citep{Vreeswijk04,Watson06}, which also resembled Galactic molecular clouds \citep{Jakobsson06b}. The first unambiguous detection of H$_2$ imprinted on a GRB afterglow spectrum was, however, first observed in GRB\,080607 at $z=3.0363$ \citep{Prochaska09}. One initial explanation for the apparent paucity of molecular hydrogen in GRB host galaxies could be that the H$_2$ molecules are photodissociated by the GRB event itself. The intense prompt $\gamma$-ray flash and afterglow emission will, however, only impact gas in the vicinity of the GRB \citep[out to 10 pc;][]{Draine02}. The cold and neutral absorbing gas is typically found to be located at much greater distances from the GRB (see e.g. Fig~\ref{fig:losprop}), which can be inferred either from the time variability of line transitions excited by the GRB afterglow flux or the minimum distance required to sustain a significant fraction of neutral gas \citep{Prochaska06a}. Due to the limited data with high enough quality it was at first difficult to determine if this was a real effect. Based on the XS-GRB afterglow legacy survey, \citet{Bolmer19} performed the first statistical analysis of the presence of H$_2$ in GRB afterglow spectra and concluded that there is actually no lack of H$_2$ in GRB host absorption systems compared to the overall population of quasar DLAs. The overall fraction of H$_2$ detections are higher in the GRB absorber population (by selection), but above the threshold of $N$(\hi) > $10^{21.7}$\,cm$^{-2}$ where the {\hi}-to-H$_2$ transition is believed to occur, the detection rates are consistent for both DLA populations.

The absorption signatures of H$_2$ are the Lyman and Werner (L-W) bands which are located bluewards of the Ly$\alpha$ absorption line, and is therefore only observable from the ground in absorbing galaxies at $z\gtrsim 2$. Due to the location of the L-W bands, high spectral resolution is required to distinguish the line transitions associated with H$_2$ from the wealth of absorption features from the Ly$\alpha$ forest. High resolution spectra are, however, only obtained for the brightest and therefore also likely less dust-obscured GRBs \citep{Ledoux09}. The sightlines most likely to exhibit H$_2$ absorption are therefore also the most difficult to observe due to the faintness of the afterglows. The absorption signatures from neutral atomic-carbon (\ci) are another efficient probe of the shielded molecular gas-phase and can be used to study the cold, neutral medium in the ISM. {\ci} is typically observed to be coincident with H$_2$ in quasar DLAs \citep[e.g.][]{Srianand05}, so targeting \ci~systems should reveal absorbers rich in molecular gas \citep{Ledoux15}. This is the basis of Chapter~\ref{chap:cisurvey}. The coincidence of \ci~with H$_2$ is likely related to the ionization potential of \ci~(11.26\,eV) being similar to the energy range of Lyman-Werner photons that can photodissociate H$_2$ (11.2 -- 13.6\,eV). The most prominent \ci\,$\lambda\lambda$\,1560,1656 line transitions are also located far from the Ly$\alpha$ forest and can be identified even in low- to medium-resolution spectroscopy.

\subsubsection*{Determining line abundances}

\begin{figure}[!t]
	\centering
	\includegraphics[width=11cm]{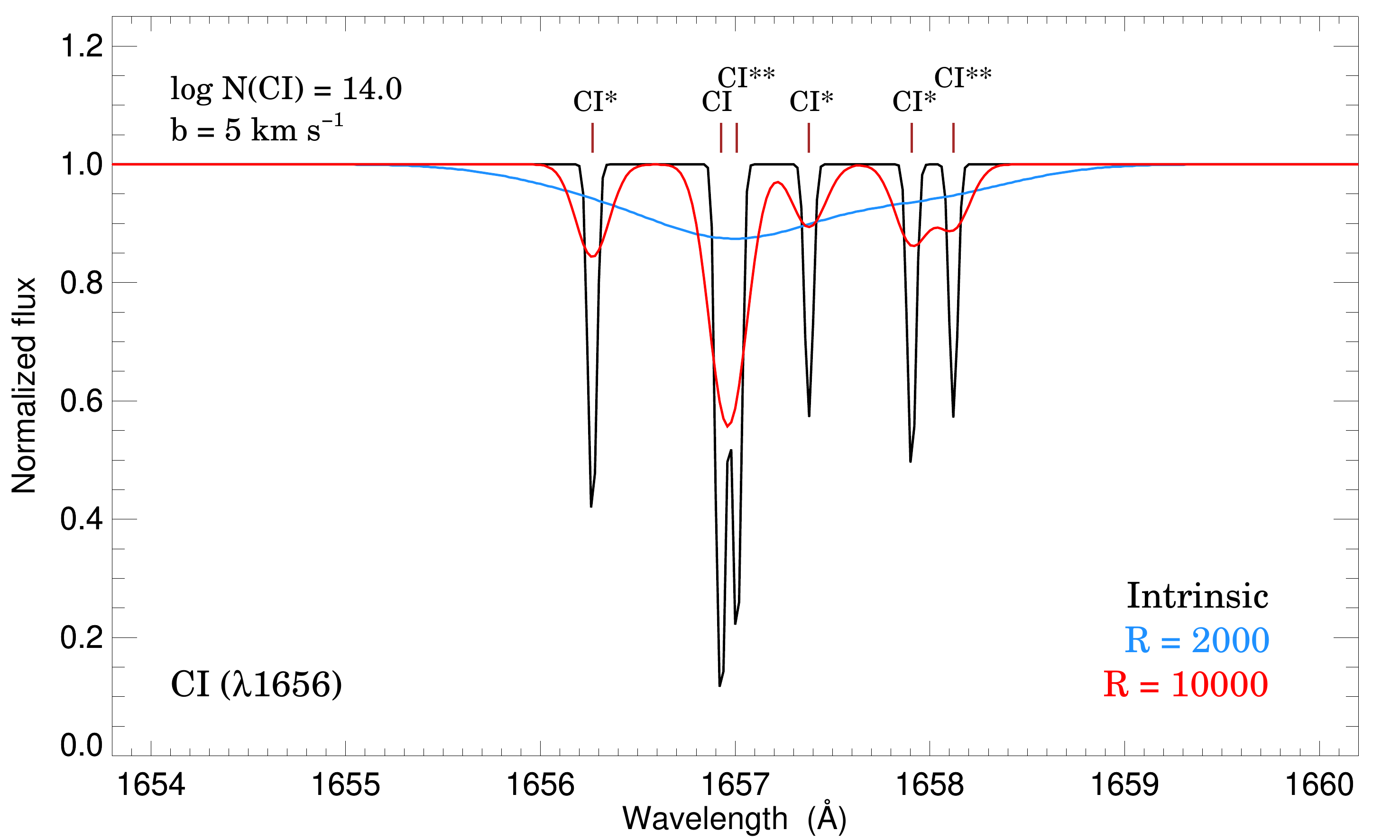}
	\caption{Synthetic absorption line profiles of the C\,{\sc i}\,$\lambda$1656 ground-state and exited transitions. The intrinsic absorption line (in black) with a total column density of $\log N$(C\,{\sc i}) = 14.0 and broadening parameter $b=5$\,km\,s$^{-1}$ is shown together with two identical lines but convolved by representative instrumental resolutions of $\mathcal{R} = 2\,000$ (in blue) and $\mathcal{R} = 10\,000$ (in red). This illustrates that in low-resolution spectroscopy, the line profiles might be affected by hidden saturation, making intrinsically satured lines appear non-saturated due to instrumental broadening. This effect also removes any informtion on the intrinsic velocity structure.}
	\label{fig:cispec} 
\end{figure}

Since the lines associated with H$_2$ and \ci~typically originate from cold gas, the intrinsic line broadening is expected to be small. The majority of GRB afterglow spectra studied in this thesis showing \ci~are observed with the X-shooter spectrograph mounted at the ESO-VLT. In Fig.~\ref{fig:cispec}, it is shown how an observed {\ci} line complex changes drastically depending on the spectral resolution. It is clear from the figure that the instrumental broadening will at some point smooth out the intrinsic profile such that all information of the relative abundances are lost. At low spectral resolution, there might also be so-called \lq hidden\rq~saturation \citep[e.g.][]{Prochaska06b}, where the lines appear to be well constrained but is in fact intrinsically saturated. The equivalent width (EW) of the lines are, however, not dependent on the resolution so the measured EWs should be the same in all cases: in the example {\ci} line complex shown in Fig.~\ref{fig:cispec}, the convolved versions of the lines are indeed measured to all have total EWs of 0.21\,\AA. If several components are blended together, however, the measured EWs will tend to underestimate the actual total element abundance. The \ci~abundances should therefore be carefully derived, which was a large part of the work done for the results presented in Chapter~\ref{chap:h2physprop}.

Typically, the line abundances are derived by modelling the observed lines with Voigt profiles. The Voigt profile is a probability distribution given by a convolution of the Gaussian and Lorentzian distribution functions. For practical purposes, an analytical approximation is typically used \citep[e.g.][]{TepperGarcia06}, which is also adopted in the line fitting routine used throughout this thesis \citep[{\tt VoigtFit};][]{Krogager18}. Other tools relying on a Bayesian approach to line fitting have also recently been developed \citep{Liang17,Bolmer19}. Common to both approaches is that they calculate the optical depth, $\tau$, for the total number of transitions for the observed transmittance of $I(\lambda) = e^{-\tau (\lambda)}$. $\tau$ is then determined by the column density of the element $X$, given a set of atomic parameters describing the line strength, $f_i$, the damping constant, $\Lambda_i$, and the resonance wavelength, $\lambda_i$, for any given transition, $i$. The fit also includes an estimate of the broadening parameter $b$, which is a combination of turbulent and Doppler broadening and therefore carries information about the velocity and temperature of the gas. A large part of the work presented in Chapters~\ref{chap:highion} and \ref{chap:h2physprop}, is based on deriving column densities for various elements using this approach. 

\vspace{1cm}

Most of the groundwork for this thesis has now been outlined, laying the foundation for the work presented in the following chapters. The topics covered in this thesis are divided into three parts, each with their own specific focus. The unifying theme, common to all three parts, are the use of GRBs and quasars as probes of dust, gas and metals in the interstellar medium of galaxies through cosmic time.

\clearpage

\thispagestyle{plain}

\vspace*{3cm}

\begin{center}
	\LARGE {\sc\underline{Note from the author}} 
\end{center}

\vspace{0.5cm}

\noindent	\normalsize In this version of the PhD thesis, the individual chapters that presents published manuscripts are not provided in full length. They are instead available on arXiv.org and in the respective journals. Here, only the title, author list and abstract is provided together with the bibliographic information. \\\\
A publication list can be found at the end of the thesis, updated on the day of arXiv submission.

\clearpage

\part{\sc The Circumburst Regions of Gamma-Ray Bursts}

\cleardoublepage

\chapter{Dust in the circumburst medium}\label{chap:140506a}

This chapter is based on the following article:

\vspace{0.5cm}

\begin{adjustwidth}{1.5em}{0pt}
	\textbf{\large Steep extinction towards GRB 140506A reconciled from host galaxy observations: Evidence that steep reddening laws are local}
\end{adjustwidth}
\vspace{0.5cm}
\noindent
Published in Astronomy \& Astrophysics, vol. 601, id. A83, 10 pp. (2017) 

\vspace{0.5cm}
\noindent
Authors: 
\begin{adjustwidth}{1.5em}{0pt}
	K. E. Heintz, J. P. U. Fynbo, P. Jakobsson, T. Kr\"uhler, L. Christensen, D. Watson, C. Ledoux, P. Noterdaeme, D. A. Perley, H. Rhodin, J. Selsing, S. Schulze, N. R. Tanvir, P. M\o ller, P. Goldoni, D. Xu \& B. Milvang-Jensen
\end{adjustwidth}

\vspace{1.5cm}


We present the spectroscopic and photometric late-time follow-up of the host
galaxy of the long-duration \textit{Swift} $\gamma$-ray burst
GRB\,140506A at $z=0.889$. The optical and near-infrared afterglow of this GRB 
had a peculiar spectral energy distribution (SED) with a strong flux-drop at 
8000\,\AA~(4000\,\AA~rest-frame) suggesting an unusually steep
extinction curve. By analysing the contribution and physical properties of the host galaxy, we 
here aim at providing additional information on the properties and origin of this 
steep, non-standard extinction. We find that the strong flux-drop in the GRB
afterglow spectrum at $< 8000$\,\AA~and rise at $< 4000$\,\AA~(observers frame)
is well explained by the combination of a steep extinction curve along the GRB
line of sight and contamination by the host galaxy light at short wavelengths
so that the scenario with an extreme 2175\,\AA~extinction bump can be excluded.
We localise the GRB to be at a projected distance of approximately 4 kpc from
the centre of the host galaxy. Based on
emission-line diagnostics of the four detected nebular lines, H$\alpha$, H$\beta$,
[O\,\textsc{ii}] and  [O\,\textsc{iii}], we find the host to be a modestly 
star forming (SFR =
$1.34\pm 0.04~M_{\odot}$ yr$^{-1}$) and relatively metal poor ($Z=0.35^{+0.15}_{-0.11}~Z_{\odot}$) 
galaxy with a large dust content, characterised by a measured
visual attenuation of $A_V=1.74\pm 0.41$ mag. We compare the host
to other GRB hosts at similar redshifts and find that it is unexceptional in
all its physical properties. We model the extinction curve of the
host-corrected afterglow and show that the standard dust properties causing the
reddening seen in the Local Group are inadequate in describing the steep drop.
We thus conclude that the steep extinction curve seen in the afterglow towards
the GRB is of exotic origin and is sightline-dependent only, further confirming
that this type of reddening is present only at very local scales and that it is
solely a consequence of the circumburst environment.

\cleardoublepage

\chapter{Probing gas and metals immediate to GRBs}\label{chap:highion}

This chapter is based on the following article:

\vspace{0.5cm}

\begin{adjustwidth}{1.5em}{0pt}
	\textbf{\large Highly ionized metals as probes of the circumburst gas in the natal regions of gamma-ray bursts}
\end{adjustwidth}
\vspace{0.5cm}
\noindent
Published in Monthly Notices of the Royal Astronomical Society, vol. 479, issue 3, 21 pp. (2018) 

\vspace{0.5cm}
\noindent
Authors: 
\begin{adjustwidth}{1.5em}{0pt}
	K. E. Heintz, D. Watson, P. Jakobsson, J. P. U. Fynbo, J. Bolmer, M. Arabsalmani, Z. Cano, S. Covino, V. D'Elia, A. Gomboc, J. Japelj, L. Kaper, J.-K. Krogager, G. Pugliese, R. S\'anchez-Ram\'irez, J. Selsing, M. Sparre, N. R. Tanvir, C. C. Th\"one, A. de Ugarte Postigo \& S. D. Vergani
\end{adjustwidth}

\vspace{1.5cm}


We present here a survey of high-ionization absorption lines in the
afterglow spectra of long-duration gamma-ray bursts (GRBs) obtained with the
VLT/X-shooter spectrograph. Our main goal is to investigate the circumburst
medium in the natal regions of GRBs. Our primary focus is on the
N\,\textsc{v}\,$\lambda\lambda$\,1238,1242 line transitions, but we
also discuss other high-ionization lines such as O\,\textsc{vi}, C\,\textsc{iv}
and Si\,\textsc{iv}. We find no correlation between the column density of
N\,\textsc{v} and the neutral gas properties such as metallicity, H\,\textsc{i}
column density and dust depletion, however the relative velocity of
N\,\textsc{v}, typically a blueshift with respect to the neutral gas, is found
to be correlated with the column density of H\,\textsc{i}. This may be
explained if the N\,\textsc{v} gas is part of an H\,\textsc{ii} region hosting
the GRB, where the region's expansion is confined by dense, neutral gas in the
GRB's host galaxy. We find tentative evidence (at $2\sigma$ significance) that
the X-ray derived column density, $N_{\mathrm{H,X}}$, may be correlated with
the column density of N\,\textsc{v}, which would indicate that both
measurements are sensitive to the column density of the gas located in the
vicinity of the GRB. We investigate the scenario where N\,\textsc{v} (and also
O\,\textsc{vi}) is produced by recombination after the corresponding atoms have
been stripped entirely of their electrons by the initial prompt emission, in
contrast to previous models where highly-ionized gas is produced by
photoionization from the GRB afterglow.

\cleardoublepage

\part{\sc Cold and Molecular Gas in High-Redshift GRB Hosts}

\cleardoublepage

\chapter{Identifying high-redshift molecular clouds}\label{chap:cisurvey}

This chapter is based on the following article:

\vspace{0.5cm}

\begin{adjustwidth}{1.5em}{0pt}
	\textbf{\large Cold gas in the early Universe. Survey for neutral atomic-carbon in GRB host galaxies at $1 < z < 6$ from optical afterglow spectroscopy}
\end{adjustwidth}
\vspace{0.5cm}
\noindent
Published in Astronomy \& Astrophysics, vol. 621, id. A20, 13 pp. (2019)

\vspace{0.5cm}
\noindent
Authors: 
\begin{adjustwidth}{1.5em}{0pt}
	K. E. Heintz, C. Ledoux, J. P. U. Fynbo, P. Jakobsson, P. Noterdaeme, J.-K. Krogager, J. Bolmer, P. M\o ller, S. D. Vergani, D. Watson, T. Zafar, A. De Cia, N. R. Tanvir, D. B. Malesani, J. Japelj, S. Covino \& L. Kaper
\end{adjustwidth}

\vspace{1.5cm}


We present a survey for neutral atomic-carbon (C\,\textsc{i}) along gamma-ray burst (GRB) sightlines, which probes the shielded neutral gas-phase in the interstellar medium (ISM) of GRB host galaxies at high redshift. We compile a sample of 29 medium- to high-resolution GRB optical afterglow spectra spanning a redshift range through most of cosmic time from $1 < z < 6$. We find that seven ($\approx 25\%$) of the GRBs entering our statistical sample have C\,\textsc{i} detected in absorption. It is evident that there is a  strong excess of cold gas in GRB hosts compared to absorbers in quasar sightlines. We investigate the dust properties of the GRB C\,\textsc{i} absorbers and find that the amount of neutral carbon is positively correlated with the visual extinction, $A_V$, and the strength of the 2175\,\AA~dust extinction feature, $A_{\mathrm{bump}}$. GRBs with C\,\textsc{i} detected in absorption are all observed above a certain threshold of $\log N$(H\,\textsc{i})$/\mathrm{cm}^{-2}$ + [X/H] > 20.7 and a dust-phase iron column density of $\log N$(Fe)$_{\mathrm{dust}}/\mathrm{cm}^{-2}$ > 16.2. In contrast to the SED-derived dust properties, the strength of the C\,\textsc{i} absorption does not correlate with the depletion-derived dust properties. This indicates that the GRB C\,\textsc{i} absorbers trace dusty systems where the dust composition is dominated by carbon-rich dust grains. The observed higher metal and dust column densities of the GRB C\,\textsc{i} absorbers compared to H$_2$- and C\,\textsc{i}-bearing quasar absorbers is mainly a consequence of how the two absorber populations are selected, but is also required in the presence of intense UV radiation fields in actively star-forming galaxies. 

\cleardoublepage

\chapter{Dust in the molecular gas-phase of the ISM}\label{chap:cidust}

This chapter is based on the following article:

\vspace{0.5cm}

\begin{adjustwidth}{1.5em}{0pt}
	\textbf{\large On the dust properties of high-redshift molecular clouds and the connection to the 2175\,{\AA} extinction bump}
\end{adjustwidth}
\vspace{0.5cm}
\noindent
Published in Monthly Notices of the Royal Astronomical Society, vol. 486, issue 2, 12 pp. (2019)

\vspace{0.5cm}
\noindent
Authors: 
\begin{adjustwidth}{1.5em}{0pt}
	K. E. Heintz, T. Zafar, A. De Cia, S. D. Vergani, P. Jakobsson, J. P. U. Fynbo, D.~Watson, J. Japelj, P. M\o ller, S. Covino, L. Kaper \& A. C. Andersen
\end{adjustwidth}

\vspace{1.5cm}


We present a study of the extinction and depletion-derived dust properties of gamma-ray burst (GRB) absorbers at $1<z<3$ showing the presence of neutral carbon (\ci). By modelling their parametric extinction laws, we discover a broad range of dust models characterizing the GRB {\ci} absorption systems. In addition to the already well-established correlation between the amount of {\ci} and visual extinction, $A_V$, we also observe a correlation with the total-to-selective reddening, $R_V$. All three quantities are also found to be connected to the presence and strength of the 2175\,\AA~dust extinction feature. 
While the amount of {\ci} is found to be correlated with the SED-derived dust properties, we do not find any evidence for a connection with the depletion-derived dust content as measured from [Zn/Fe] and $N$(Fe)$_{\rm dust}$. 
To reconcile this, we discuss a scenario where the observed extinction is dominated by the composition of dust particles confined in the molecular gas-phase of the ISM. We argue that since the depletion level trace non-carbonaceous dust in the ISM, the observed extinction in GRB {\ci} absorbers is primarily produced by carbon-rich dust in the molecular cloud and is therefore only observable in the extinction curves and not in the depletion patterns. This also indicates that the 2175\,\AA~dust extinction feature is caused by dust and molecules in the cold and molecular gas-phase. This scenario provides a possible resolution to the discrepancy between the depletion- and SED-derived amounts of dust in high-$z$ absorbers.

\cleardoublepage

\chapter{Physical properties of molecular clouds in GRB hosts}\label{chap:h2physprop}

This chapter is based on the following article:

\vspace{0.5cm}

\begin{adjustwidth}{1.5em}{0pt}
	\textbf{\large New constraints on the physical conditions in H$_2$-bearing GRB-host damped Lyman-$\alpha$ absorbers}
\end{adjustwidth}
\vspace{0.5cm}
\noindent
Published in Astronomy \& Astrophysics, vol. 629, id. A131, 21 pp. (2019)

\vspace{0.5cm}
\noindent
Authors: 
\begin{adjustwidth}{1.5em}{0pt}
	K. E. Heintz, J. Bolmer, C. Ledoux, P. Noterdaeme, J.-K. Krogager, J. P. U. Fynbo, P. Jakobsson, S. Covino, V. D'Elia, M. De Pasquale, D. H. Hartmann, L. Izzo, J. Japelj, D. A. Kann, L. Kaper, P. Petitjean, A. Rossi, R. Salvaterra, P. Schady, J. Selsing, R. Starling, N. R. Tanvir, C. C. Th\"one,
	A. de Ugarte Postigo, S. D. Vergani, D. Watson, K. Wiersema \& T. Zafar
\end{adjustwidth}

\vspace{1.5cm}


We report the detections of molecular hydrogen (H$_2$), vibrationally-excited H$_2$ (H$^*_2$), and neutral atomic carbon (\ci), an efficient tracer of molecular gas, in two new afterglow spectra of GRBs\,181020A ($z=2.938$) and 190114A ($z=3.376$), observed with X-shooter at the Very Large Telescope (VLT). Both host-galaxy absorption systems are characterized by strong damped Lyman-$\alpha$ absorbers (DLAs) and substantial amounts of molecular hydrogen with $\log N$(\hi, H$_2$) = $22.20\pm 0.05,~20.40\pm 0.04$ (GRB\,181020A) and $\log N$(\hi, H$_2$) = $22.15\pm 0.05,~19.44\pm 0.04$ (GRB\,190114A). The DLA metallicites, depletion levels, and dust extinctions are within the typical regimes probed by GRBs with [Zn/H] = $-1.57\pm 0.06$, [Zn/Fe] = $0.67\pm 0.03$, and $A_V = 0.27\pm 0.02$\,mag (GRB\,181020A) and [Zn/H] = $-1.23\pm 0.07$, [Zn/Fe] = $1.06\pm 0.08$, and $A_V = 0.36\pm 0.02$\,mag (GRB\,190114A). In addition, we examine the molecular gas content of all known H$_2$-bearing GRB-DLAs and explore the physical conditions and characteristics required to simultaneously probe {\ci} and H$^*_2$. We confirm that H$_2$ is detected in all {\ci}- and H$^*_2$-bearing GRB absorption systems, but that these rarer features are not necessarily detected in all GRB H$_2$ absorbers. We find that a large molecular fraction of $f_{\rm H_2} \gtrsim 10^{-3}$ is required for {\ci} to be detected. The defining characteristic for H$^*_2$ to be present is less clear, though a large H$_2$ column density is an essential factor. We find that the observed line profiles of the molecular-gas tracers are kinematically \lq\lq cold\rq\rq, with small velocity offsets of $\delta v < 20$\kms~from the bulk of the neutral absorbing gas. We then derive the H$_2$ excitation temperatures of the molecular gas and find that they are relatively low with $T_{\rm ex} \approx 100 - 300$\,K, however, there could be evidence of warmer components populating the high-$J$ H$_2$ levels in GRBs\,181020A and 190114A. Finally, we demonstrate that even though the X-shooter GRB afterglow campaign has been successful in recovering several H$_2$-bearing GRB-host absorbers, this sample is still hampered by a significant dust bias excluding the most dust-obscured H$_2$ absorbers from identification. {\ci} and H$^*_2$ could open a potential route to identify molecular gas even in low-metallicity or highly dust-obscured bursts, though they are only efficient tracers for the most H$_2$-rich GRB-host absorption systems. 

\cleardoublepage

\part{\sc Quasars Concealed by Dusty and Metal-Rich Absorbers}

\cleardoublepage





\chapter{Selecting dust-reddened quasars}\label{chap:kvrq}








There are at least two ways to gauge the extent of the proposed dust bias in quasar-selected damped Lyman-$\alpha$ absorber (DLA) samples. One is to design a selection that is specifically tailored for the identification of the most dust-reddened systems, that have been missed in current surveys. The focus of this Chapter will be to define a set of selection criteria, specifically targeting quasars at $z \gtrsim 2$ with intervening dusty DLAs based on their photometry that are not identified in classical quasar surveys. In Chapter~\ref{chap:dustydla}, a prime example of one such system is presented, identified via these new selection criteria. 

Another approach to assess the true underlying population of quasars is to define a complete sample. Here, all quasar candidates should be selected uniformly, dependent only on their most basic properties. For such an approach to be effective, this should also include a rejection of the vast majority of local stellar sources, which is typically done in color-space. This might, however, bias the sample against quasars that photometrically resemble stars the most. Ideally, all quasar candidates should be selected regardless of their intrinsic spectral properties within the limiting magnitude of the survey. Such a selection is presented in Chapter~\ref{chap:gaiasel}, based on the astrometric measurements from the {\it Gaia} mission.

\section{Photometric data}

\subsection{KiDS and VIKING survey data}

\begin{figure} [!t]
	\epsfig{file=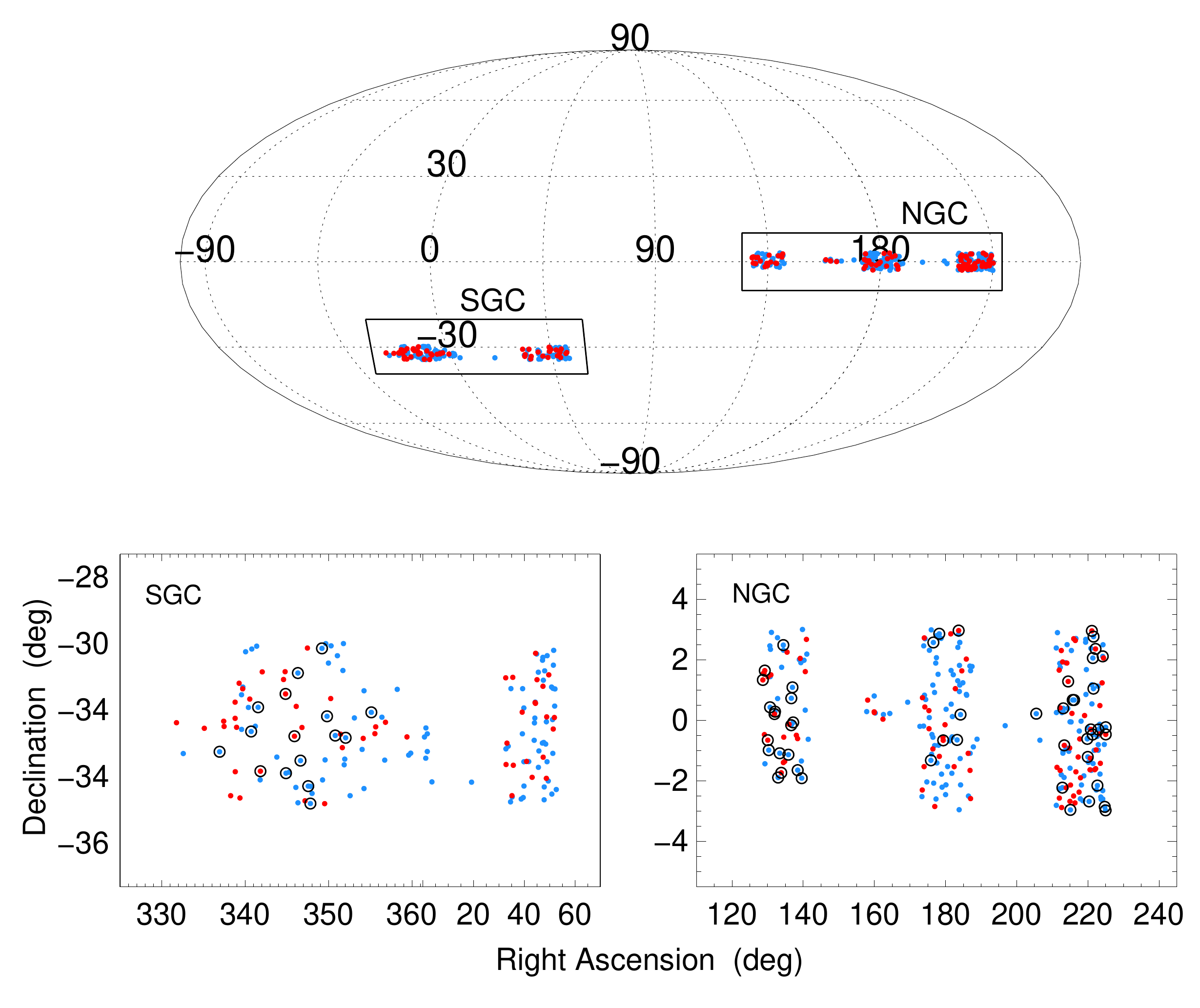,width=12cm}
	\caption{At the top is shown a Mollweide projection of the positions on the sky of the full list of quasar candidates from the red and blue samples (color-coded accordingly). In the two bottom panels are shown zoom-ins on the two sample regions, the NGC and SGC. The quasar candidates from the red and blue sample are here color-coded as above, where the circles denote quasar candidates for which follow-up spectroscopy was obtained as part of the KV-RQ survey presented here.
	}
	\label{fig:skycov}
\end{figure}

The photometric catalag on which the selection presented here is based, was compiled by cross-matching the Kilo-Degree Survey Data Release 3 \citep[KiDS-DR3;][]{deJong17}, the VISTA Kilo-Degree INfrared Galaxy Data Release 2 \citep[VIKING-DR2;][]{Edge13,Edge16} survey, and the \textit{WISE} AllSKY data release \citep[AllWISE;][]{Wright10,Cutri13}. The KiDS and VIKING are so-called sister surveys in the sense that they aim to cover the same $\sim 1500$ deg$^2$ on the sky. The survey areas are divided into two main strips, one in the Northern and one in the Southern Galactic Cap (NGC and SGC). The NGC is centred on $(\alpha,\delta)_{\mathrm{NGC}}=(12~\mathrm{h},~0^{\mathrm{o}})$ and the SGC is centred on $(\alpha,\delta)_{\mathrm{SGC}}=(0.8~\mathrm{h},~-30^{\mathrm{o}})$. The coverage of KiDS-DR3 and VIKING-DR2 is so far only a subset of the aimed 1500 deg$^2$. The current cross-match between the two surveys covers $\approx 340$ deg$^2$ in the NGC and $\approx 225$ deg$^2$ in the SGC (total of $\approx 565$ deg$^2$, see e.g. Fig.~\ref{fig:skycov}).

KiDS covers the optical Sloan \textit{u, g, r} and $i$ filters at 3550, 4775, 6230 and 7360\,\AA, respectively, using the OmegaCAM wide-field imager \citep{Kuijken11} on the VLT Survey Telescope (VST). The VIKING survey covers the near-infrared \textit{Z, Y, J, H} and $K_S$ filters at 8770, 10\,200, 12\,520, 16\,450 and 21\,470\,\AA, respectively, using the VISTA InfraRed CAMera \citep[VIRCAM;][]{Dalton06} at the VISTA telescope \citep{Emerson06}. In the public available KiDS catalog, only fluxes for each source were listed in the relevant filters scaled such that the AB zero point is zero. The conversion from flux, $f$, into AB magnitudes is therefore $m_{\rm AB} = -2.5\log_{10}(f)$. To obtain the associated errors in the AB magnitude system, $\sigma_{m}$, the errors on the flux, $\sigma_{f}$, can be propagated as $\sigma_{m} =  2.5\log_{10}(e) \times \sigma_{f} / f$. The magnitudes for each source are included in the VIKING catalog, but given in the Vega magnitude system. To convert these to AB magnitudes, the following conversion factors have been applied\footnote{From the Cambridge Astronomy Survey Unit (CASU): \url{http://casu.ast.cam.ac.uk/surveys-projects/vista/technical/filter-set}} 
\begin{align*}
Z_{\rm AB} =&~Z_{\rm Vega} + 0.521, ~Y_{\rm AB} = Y_{\rm Vega} + 0.618, ~J_{\rm AB} = J_{\rm Vega} +  0.937, \\ 
&H_{\rm AB} = H_{\rm Vega} + 1.384 ~{\rm and}~ K_{S,\rm AB} = K_{S,\rm Vega} + 1.839
\end{align*}
All magnitudes from both the KiDS and VIKING catalog used here are obtained with an aperture size of $2\farcs 0$. The KiDS and VIKING surveys are $\sim 2$ mag deeper than the SDSS and UKIDSS surveys, and are thus an ideal extension to previous searches for faint quasars strongly reddened by dusty DLAs. The AllWISE survey covers the mid-infrared filters $W_1$, $W_2$, $W_3$ and $W_4$, with effective peak wavelengths at 3.4, 4.6, 12 and 22 $\mu$m, respectively. The magnitudes in this catalog is also given in the Vega system, but these are not converted in the final catalog.

\subsection{Combining the survey catalogs}

The final, combined photometric catalog was constructed by B. Milvang-Jensen, by first stitching the 440 (KiDS) and 477 (VIKING) tiles together, to produce a single KiDS+VIKING catalog. For the latter, it was required that the VIKING catalog entry \texttt{PRIMARY\_SORUCE} = 1 to remove duplicates due to overlapping sky coverage of the individual VIKING tiles. The KiDS and VIKING catalogs were then cross-matched by requiring that; 1) the source should be detected both in KiDS and VIKING, 2) only the closest, best-match counterpart is saved and 3) VIKING sources should be at a maximum distance of $1\farcs0$ from the corresponding KiDS source. The typical separation distance between the optical and near-infared counterparts is $0\farcs12$. The AllWISE catalog was then cross-matched to this combined catalog, including only the best-match counterpart within $2\farcs0$. Only about one-third of the roughly $1.5\times 10^7$ entries in the combined KiDS-VIKING catalog were found to have a counterpart in the AllWISE survey within $2\farcs0$. This is most likely related to the shallower survey depth of the {\it WISE} mission. A large fraction ($\sim 95\%$) of the sources in the KiDS-VIKING catalog without {\it WISE} detections also show optical/near-infrared colors consistent with main-sequence stars. These are therefore less likely to be detected by {\it WISE} due to their weak mid-infrared emission.

\section{Target selection}

The present selection of the candidate quasars is based on a combination of optical to near/mid-infrared photometry. The candidates were required to be detected in (at least) the \textit{g, r, i, J} and $K_S$ filters and in the first three \textit{WISE} bands with a signal-to-noise ratio of at least 3 (i.e., S/N$_{W1-3}\geq 3$). Furthermore, the candidates should be flagged as point sources in both KiDS and VIKING. 
In the KiDS catalog a morphology flag is given for each source where \texttt{SG2DPHOT} = 1 (for high-confidence stars), 2 (objects with a FWHM less than the typical stellar locus), 4 (stars following their star/galaxy classification) and 0 (for likely galaxies). These values are additive, so a source with \texttt{SG2DPHOT} = 5 is a high-confidence star which also fulfills their star/galaxy classification. Only sources with \texttt{SG2DPHOT} $\ge 1$ were included in the final sample. For the VIKING data, it was required that \texttt{pGal} < 0.90 \citep[similar to, but more conservative than the \texttt{pGal} < 0.95 criterion used by][who also gives a detailed discussion of this cut as a star/galaxy separator]{Venemans13,Venemans15} and that \texttt{mclass} = -1,-2 where \texttt{mclass} = 1 (galaxy), 0 (noise), -1 (star), -2 (probably a star), 3 (probably a galaxy) and -9 (saturated). Finally, the candidates should be brighter than $J<20$ mag to allow for spectroscopic follow-up. The magnitude limit was defined in the near-infared to avoid biasing the sample against sources with heavily reddened optical colors.

\begin{figure} [!t]
	\centering
	\epsfig{file=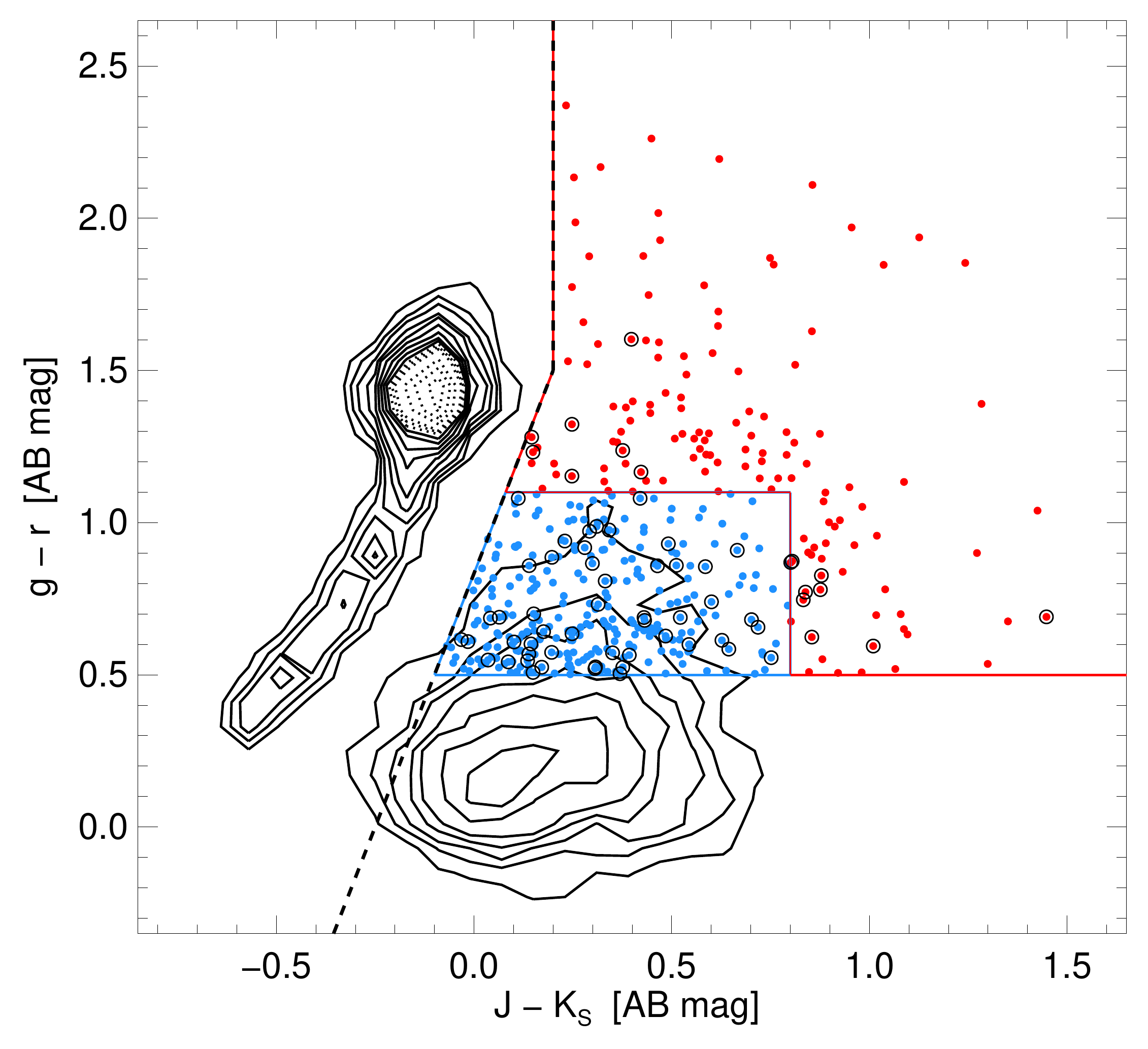,width=10cm}
	\caption{Optical and near-infrared color-color plot of the point sources brighter than $J<20$\,mag (contours) from the full combined KiDS-VIKING-AllWISE catalog. The high-likelihood quasar candidates from the red and blue samples are shown by the colored dots. The circles again denote quasars for which follow-up spectroscopy was obtained as part of the KV-RQ survey presented here. The red and blue solid lines indicate the selection criteria for the two samples, respectively. The general stellar rejection line is marked by the black dashed line.
	}
	\label{fig:stelrej}
\end{figure}

\subsection{Stellar rejection}

To increase the quasar selection efficiency it is important exclude the vast population of especially dwarf stars that most resemble quasars at optical wavelengths. Following a similar approach as the \lq KX\rq~selection \citep[e.g.][]{Warren00,Maddox08,Maddox12} to identify quasars, additional stellar rejection criteria based on the targets $J-K_S$ and $g-r$ colors were imposed (see Fig.~\ref{fig:stelrej}). These specific selection criteria were motivated by the similar cuts used in previous searches for dust-reddened quasars \citep{Fynbo13a,Krogager15,Krogager16b}. Out of the total 11,637 point sources brighter than $J<20$ mag in the combined catalog, 6601 (56.72\%) are located rightwards of the stellar rejection line shown in Fig.~\ref{fig:stelrej}. To limit the sample to only include the reddest most sources in the optical, additional color criteria of $g-r>0.5$ and $r-i>0.0$ were imposed. This leaves 1256 candidate reddened quasars, which thus constitute $\approx 20\%$ of the full, high-likelihood quasar sample.

\subsection{Limiting contamination with a red and blue sample}

To increase the selection efficiency and further reject contaminating stellar sources, the parent sample was divided into two; a blue and a red sample (again, see Fig.~\ref{fig:stelrej}). The blue sample includes all the high-likelihood quasar candidates that have $J-K_S < 0.8$ and $0.5 < g-r < 1.1$. These particular color cuts were imposed to be more inclusive with the {\it WISE} photometric cuts (described below), since quasars are well-separated from the stellar locus in this region of optical/near-infrared color-color space. The red sample was limited to only contain sources with $J-K_S > 0.8$ or $g-r > 1.1$. In this region, the {\it WISE} color criteria need to be more conservative to remove the significant population of dwarf stars that overlap with the quasar colors in the optical and near-infrared.

\subsection{High-$z$ identification with \textit{WISE} }

\begin{figure} [!t]
	\centering
	\epsfig{file=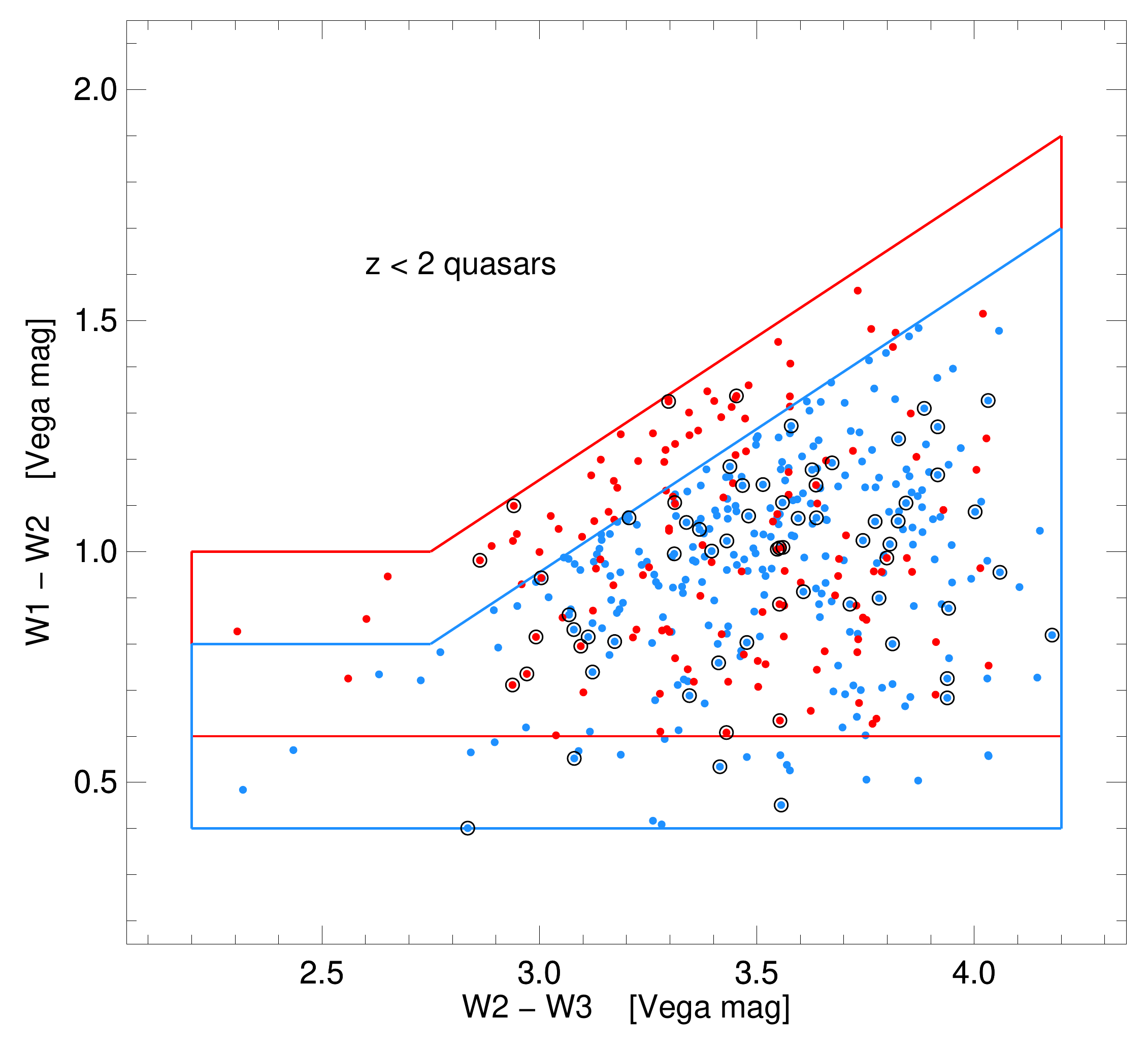,width=10cm}
	\caption{{\it WISE} color-color plot of the high-likelihood quasar candidates from the red and blue samples (color-coded accordingly). The circles again denote quasars for which follow-up spectroscopy was obtained as part of the KV-RQ survey presented here. The red and blue solid lines indicate the selection criteria for the two samples, respectively. The exclusion zone for low-redshift quasars in the top-left is highlighted in the figure as well.
	}
	\label{fig:wisesel}
\end{figure}

To specifically target quasars with intervening DLAs, it is important to constrain the selection to quasars located $z>2$ such that the Ly$\alpha$ line is observable from the ground. \cite{Krogager16b} found that most quasars at $z < 2$ can be removed based on their observed mid-infrared $W_2-W_3$ vs. $W_1-W_2$ colors. Using a similar, but more conservative, set of mid-infrared photometric color criteria (shown in Fig.~\ref{fig:wisesel}) allows a more robust rejection of quasars at $z<2$, while also being more sensitive to quasars at $z > 3$. Specifically, the {\it WISE} color cuts were defined as
\begin{align}
\mathrm{for}~&W2-W3 < 2.75: W1-W2 < 0.8 \\
\mathrm{for}~&W2-W3 > 2.75: W1-W2 < 0.62 \times (W2-W3)-0.91~. \nonumber
\end{align}
for the blue sample and
\begin{align}
\mathrm{for}~&W2-W3 < 2.75: W1-W2 < 1.0 \\
\mathrm{for}~&W2-W3 > 2.75: W1-W2 < 0.62 \times (W2-W3)-0.71~, \nonumber
\end{align}
for the red sample. The lower boundaries of $W_1-W_2 > 0.6$ (red) and $W1-W2 > 0.4$ (blue) were imposed to limit contamination from stars and galaxies while still recovering a part of the high-redshift quasars. Since the blue sample only contains a small fraction of contaminating sources, the lower boundary in $W_1 - W_2$ is less conservative such that a higher fraction of $z \gtrsim 3$ quasars are included \citep{Stern12}.

\section{Sample properties}

The goal of the red sample is to recover quasars even more reddened by dust in intervening DLAs than those found in the previous searches. The blue sample is mainly constructed to have higher survey efficiency and contain brighter candidates, but still with the potential of revealing a few, but likely less, dust-rich DLAs. In the two constructed catalogs, 139 candidate quasars are in the red sample and 280 are in the blue sample. The SDSS/BOSS quasar surveys cover only the NGC region of the two catalogs, which is also best visible from the Northern hemisphere (e.g. for observations with the Nordic Optical Telescope (NOT)). Before performing follow-up spectroscopy of any of our candidates, sources that had already been observed spectroscopically as part of SDSS or BOSS (DR13) or by us as part of the HAQ \citep{Fynbo13a,Krogager15} or eHAQ \citep{Krogager16b} surveys were removed from the target list.  
In the blue NGC sample 66 (37\%) of the candidates were already observed by SDSS/BOSS (where only 13 of these were at $z<2$, all showing a substantial amount of reddening). The remaining 53 quasars were all in the target redshift range at $2.0 < z < 3.7$. To be conservative, sources that were classified as galaxies in the SDSS/BOSS were also removed, but there were only six (3\%) such cases. Furthermore, one already observed HAQ (HAQ\,1411-0104 at $z=3.50$) and one source classified as a star in SDSS was removed as well. This leaves 106 candidate dust-reddened quasars in the blue NGC sample.
In the red NGC sample only 12 (14\%) of the candidates were already observed by SDSS/BOSS (where only two of these were at $z<2$, both showing a substantial amount of reddening). The remaining 10 quasars were all in the target redshift range at $2.0 < z < 4.3$. Again, sources that were classified as galaxies in the SDSS/BOSS were removed, which were only relevant for seven (8\%) such cases in this redder sample. Furthermore, one already observed eHAQ (eHAQ\,1136+0027 at $z=1.12$) and one source classified as a star in SDSS were removed as well. This leaves 65 candidate dust-reddened quasars in the red NGC sample. It is clear that the SDSS/BOSS selection algorithms are much less efficient at identifying the reddest most sources in the optical, as represented by the recovering fraction of the red vs. blue sample.

\begin{figure} [!t]
	\epsfig{file=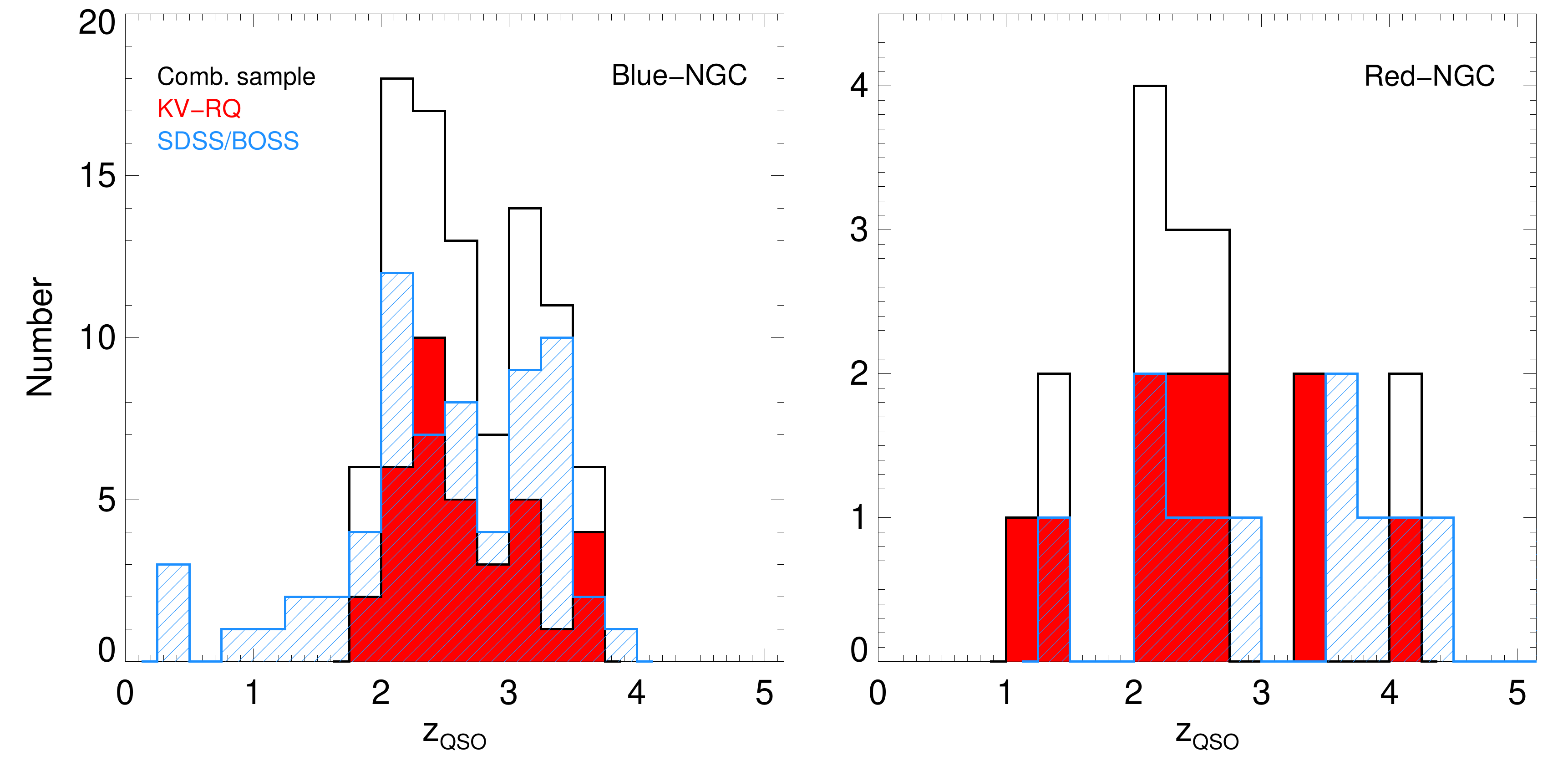,width=12cm}
	\caption{Quasar redshift distribution of the red and blue NGC samples. The filled red histograms show the $z_{\rm QSO}$ distribution of the quasars identified as part of the KV-RQ survey. The line-filled blue histograms show the candidates from the red and blue samples that had already been observed spectroscopically as part of the SDSS or BOSS (DR13) quasar surveys. The solid black lines show the combined redshift distributions.
	}
	\label{fig:zplot}
\end{figure}

In Fig.~\ref{fig:zplot} is shown the redshift distribution of the quasars observed so far as part of the KiDS-VIKING Red Quasar (KV-RQ) survey presented here. It is evident that the KV-RQ survey probes roughly the same redshift distribution as the SDSS/BOSS quasar surveys. We have, however, been able to recover more dust-reddened systems. In addition, 11 new quasars with intervening DLAs have been discovered in the blue sample so far, but unfortunately none in the red sample. One of these quasar DLAs, KV-RQ\,1500-0013, will be presented in the next Chapter~\ref{chap:dustydla}, where the general population of dusty DLAs will also be discussed.

\cleardoublepage

\chapter{Characterizing the population of dusty DLAs}\label{chap:dustydla}

This chapter is based on the following article:

\vspace{0.5cm}

\begin{adjustwidth}{1.5em}{0pt}
	\textbf{\large A quasar hiding behind two dusty absorbers. Quantifying the selection bias of metal-rich, damped Ly$\alpha$ absorption systems}
\end{adjustwidth}
\vspace{0.5cm}
\noindent
Published in Astronomy \& Astrophysics, vol. 615, id. A43, 14 pp. (2018)

\vspace{0.5cm}
\noindent
Authors: 
\begin{adjustwidth}{1.5em}{0pt}
	K. E. Heintz, J. P. U. Fynbo, C. Ledoux, P. Jakobsson, P. M\o ller, L. Christensen, S.~Geier, J.-K. Krogager \& P. Noterdaeme
\end{adjustwidth}

\vspace{1.5cm}


The cosmic chemical enrichment as measured from damped Ly$\alpha$ absorbers (DLAs) will be underestimated if dusty and metal-rich absorbers have evaded identification. Here we report the discovery and present the spectroscopic observations of a quasar, KV-RQ\,1500-0031, at $z=2.520$ reddened by a likely dusty DLA at $z=2.428$ and a strong Mg\,\textsc{ii} absorber at $z=1.603$. This quasar was identified as part of the KiDS-VIKING Red Quasar (KV-RQ) survey, specifically aimed at targeting dusty absorbers which may cause the background quasars to escape the optical selection of e.g. the Sloan Digital Sky Survey (SDSS) quasar sample. For the DLA we find an H\,\textsc{i} column density of $\log N$(H\,\textsc{i}) = $21.2\pm 0.1$ and a metallicity of [X/H] = $-0.90\pm 0.20$ derived from an empirical relation based on the equivalent width of Si\,\textsc{ii}\,$\lambda$\,1526. We observe a total visual extinction of $A_V=0.16$ mag induced by both absorbers. To put this case into context we compile a sample of 17 additional dusty ($A_V > 0.1$ mag) DLAs toward quasars (QSO-DLAs) from the literature for which we characterize the overall properties, specifically in terms of H\,\textsc{i} column density, metallicity and dust properties. From this sample we also estimate a correction factor to the overall DLA metallicity budget as a function of the fractional contribution of dusty QSO-DLAs to the bulk of the known QSO-DLA population. We demonstrate that the dusty QSO-DLAs have high metal column densities ($\log N$(H\,\textsc{i}) + [X/H]) and are more similar to gamma-ray burst (GRB)-selected DLAs (GRB-DLAs) than regular QSO-DLAs. We evaluate the effect of dust reddening in DLAs as well as illustrate how the induced color excess of the underlying quasars can be significant (up to $\sim 1$ mag in various optical bands), even for low to moderate extinction values ($A_V \lesssim 0.6$ mag). Finally we discuss the direct and indirect implications of a significant dust bias in both QSO- and GRB-DLA samples.

\cleardoublepage

\chapter{Astrometric selection of quasars}\label{chap:gaiasel}

This chapter is based on the following article:

\vspace{0.5cm}

\begin{adjustwidth}{1.5em}{0pt}
	\textbf{\large Unidentified quasars among stationary objects from Gaia DR2}
\end{adjustwidth}
\vspace{0.5cm}
\noindent
Published in Astronomy \& Astrophysics, vol. 615, id. L8, 9 pp. (2018) 

\vspace{0.5cm}
\noindent
Authors: 
\begin{adjustwidth}{1.5em}{0pt}
	K. E. Heintz, J. P. U. Fynbo, E. H\o g, P. M\o ller, J.-K. Krogager, S. Geier, P. Jakobsson \& L. Christensen
\end{adjustwidth}

\vspace{1.5cm}


Here we apply a technique selecting quasar candidates purely as sources
with zero proper motions in the \textit{Gaia} data release 2 (DR2). We
demonstrate that
this approach is highly efficient toward high Galactic latitudes with $\lesssim
25\%$ contamination from stellar sources. Such a selection technique offers
very pure sample completeness, since all cosmological point sources are
selected regardless of their intrinsic spectral properties within the limiting
magnitude of \textit{Gaia}. We carry out a pilot-study, defining a sample compiled 
by including all \textit{Gaia}-DR2 sources within one degree of the north
Galactic pole (NGP) selected to have proper motions consistent with zero within
$2\sigma$ uncertainty. By cross-matching the sample to the optical Sloan
Digital Sky Survey (SDSS) and the mid-infrared (MIR) AllWISE photometric catalogues,
we investigate the colours of each of our sources. We  determine the
efficiency of our selection by comparison with previously
spectroscopically confirmed quasars. The majority of the zero-proper-motion sources selected here
have optical to MIR colours consistent with known quasars. The remaining population
may be contaminating stellar sources, but some may also be quasars with colours
similar to stars. Spectroscopic follow-up of the zero-proper-motion sources is
needed to unveil such a hitherto hidden quasar population. This approach has
the potential to allow substantial progress on many important questions
concerning quasars, such as determining the fraction of dust-obscured quasars,
the fraction of broad absorption line (BAL) quasars, and the metallicity
distribution of damped Lyman-$\alpha$ absorbers. This technique could also
potentially reveal new types of quasars or even new classes of cosmological
point sources.

\cleardoublepage





\chapter{Summary and Outlook} \label{chap:conc}

Here I will provide a summary of the work done in this thesis and some near-future prospects that would be natural directions to continue in, based on both personal and general knowledge that has been obtained as part of this work.  

\vspace{0.2cm}

In the first part of this thesis I studied the gas, dust and metals in the circumburst regions of GRBs. In Chapter~\ref{chap:140506a}, I presented the analysis of a single burst, GRB\,140506A at $z=0.889$, and the very peculiar dust extinction curve observed in the first VLT/X-shooter afterglow spectrum of this GRB. Based on late-time follow-up observations of the host galaxy, it was possible to exclude an extreme 2175\,\AA~extinction bump as the origin of the steep extinction. Instead, we showed that the shape of the afterglow was more likely a combination of a steep extinction curve along the GRB line of sight and contamination by the host galaxy light at short wavelengths. We also derived the properties of the host galaxy and found that it was consistent with other known GRB hosts at similar redshifts and showed no sign of a steep extinction curve component, similar to that observed in the afterglow spectrum. In Chapter~\ref{chap:highion}, I presented a study of the high-ionization metal lines in a large sample of GRB absorbers observed as part of the VLT/X-shooter GRB afterglow legacy survey \citep{Selsing19}. In this work, we proposed a scenario where N\,{\sc v} is produced by recombination of atoms in the nearby gas after the nitrogen atoms had been completely ionized and stripped of their electrons by the intense flux from the GRB and following afterglow. We then explored the kinematics of the high-ionization lines compared to those observed in the singly-ionized neutral gas and found that N\,{\sc v} is likely confined to the same cloud component as the bulk of the neutral gas for large H\,{\sc i} column densities of $N$(\hi) > $10^{21.5}$\,cm$^{-2}$. The strongest evidence for N\,{\sc v} to originate in the circumburst came from the tentative correlation with the X-ray derived equivalent hydrogen column density, $N_{\rm H,X}$, that have been claimed to originate within $\sim 10$\,pc from the GRB explosition site \citep{Watson13}.  

In both of the studies outlined above, the evidence for the observed tracers of gas, dust and metals to originate in the circumburst region of the GRB is primarily based on indirect observations. One way to test (or exclude) if for example the N\,{\sc v} absorption line is produced by the GRB would be to obtain time-resolved afterglow spectra of GRB absorbers where this feature is present. The time-variability (if any) of the N\,{\sc v} abundance would then hint at the origin of this absorption feature. Another powerful way to study the GRB progenitors and the gas in the immediate region of the GRB explosion site would be through spatially resolved spectroscopic observations of the host galaxies of low-redshift GRBs using sensitive integral field unit (IFU) instruments. Especially the new sensitive Multi-Unit Spectrograph Explorer (MUSE) on the VLT will revolutionize our understanding of GRB host galaxies and their spatially resolved environments.

\vspace{0.2cm}

In the work presented in the second part of this thesis in Chapters~\ref{chap:cisurvey} through \ref{chap:h2physprop}, we demonstrated how GRBs can be used as probes of the molecular gas-phase in the ISM of high-redshift galaxies. In particular, we surveyed neutral atomic-carbon {\ci}, which is used as a tracer for molecular hydrogen H$_2$, in a large sample of GRB afterglow spectra. We found that there is a significant excess of cold gas in GRB host absorption systems, with a detection probability of $\sim 25\%$ for {\ci}, compared to $\sim 1\%$ in quasar absorbers \citep{Ledoux15}. Moreover, we found that a larger {\hi} column density and metallicity is required for GRB absorbers to show {\ci} and H$_2$ than what is typically observed in quasar absorbers, likely related to enhanced star-formation and more intense interstellar UV radiation fields in the GRB host absorption systems. We also showed that the 2175\,\AA~dust extinction feature is only detected in the most dust-reddened and {\ci}-rich absorption systems and likely produced from carriers located in the same {\ci}-bearing molecular cloud. We then characterized the physical properties of the diffuse molecular gas-phase in the ISM of H$_2$-bearing GRB host galaxy absorption systems and examined the defining characteristics for these systems to simultanously probe other molecular gas tracers such as {\ci} and vibrationally-excited H$_2$ (H$^*_2$). A large H$_2$ abundance and molecular gas fraction is found to be essential for GRB host absorbers to also show the presence of {\ci} and H$^*_2$, limiting their overall efficiency of tracing diffuse molecular gas. However, both {\ci} and H$^*_2$ could open a potential route to identify molecular clouds even in low-redshift, highly dust-obscured GRB afterglows observed with low-resolution spectroscopy. 

Now that a large sample of GRB and quasar absorption systems with molecular gas has been detected at high-$z$, the next natural step is to examine how the properties of the cold and molecular gas derived from absorption and emission are connected. One way to do this is by targeting the CO emission of these systems and compare it to the molecular gas tracers observed in absorption \citep{Neeleman16}. I have also recently applied for a similar project to search for CO emission in a sample of H$_2$-rich GRB host galaxy absorbers using the Atacama Large Millimeter/submillimeter Array (ALMA). These studies would allow us to couple the wealth of information on the molecular gas derived in absorption to that observed in emission. Moreover, based on preliminary results (Heintz \& Watson, in preparation), it appears that it is actually possible to use GRB and quasar absorbers to obtain a direct calibration of the {\ci} (and CO) luminosity to the molecular gas mass of high-$z$ galaxies, which would be extremely beneficial to the extensive samples of emission-selected galaxies for which molecular line emission has been detected.

\vspace{0.2cm}

The focus of the final part of this thesis was to understand and explore the possible effects quasar selection bias had on the census of neutral gas and metals in high-$z$ absorption-selected galaxies. One way to gauge this effect is to design selection criteria that are tailored to identify dusty and metal-rich absorbers that have otherwise been missed in typical quasar surveys. In Chapter~\ref{chap:kvrq}, I presented such a selection that were found to be very efficient in identifying $z>2$ dust-reddened quasars. In the work done for Chapter~\ref{chap:dustydla}, we characterize one of the foreground dusty damped Lyman-$\alpha$ absorbers (DLAs) found in one such sightline. The DLA was found to have $\sim 10\%$ solar metallicity and visual extinction of $A_V = 0.16$\,mag, demonstrating that even absorption systems with modest metal and dust content can cause the background quasar to drop out of typical optical selection criteria. We then compiled a sample of DLAs with similar dust content and showed that this missing population of quasar DLAs better resemble absorption systems observed in GRB sightlines. The tension in the chemical enrichment as a function of redshift of the GRB absorbers and the typical quasar DLA population would therefore be relieved if these dusty quasar DLAs are properly taken into account. Another way to assess the full underlying quasar population would be to define a new, less biased way of identifying quasars. In the work presented in Chapter~\ref{chap:gaiasel}, we demonstrated that it is possible to efficiently select quasars at high Galactic latitudes based on their zero proper motion on the sky using the astrometric data from the {\it Gaia} satellite. Specifically, we found that the relative number of quasars to stars within one degree of the North Galactic Pole was $\sim 75\%$ based on the photometry of the zero-proper-motion sources.
 
Now that we have demonstrated that there is indeed a dust bias against the most metal-rich DLAs, the next crucial step is to quantify this effect more robustly. In the recent work by \citet{Krogager19}, we found that the dust bias arises as an effect of the magnitude and colour criteria utilized in the Sloan Digital Sky Survey (SDSS) quasar target selection. As a consequence, DLAs identified as part of this survey would underestimate the mass density of neutral hydrogen by 10 to 50\%, and the mass density in metals by at least 30\% at $z\sim 3$. For the {\it Gaia}-selection of quasars, the next step would be to obtain spectroscopy of a complete sample of zero-proper-motion sources to quantify the efficiency directly. We currently have one accepted proposal at the Nordic Optical Telescope to observe such a sample. 
This way of identifying quasars, based solely on zero proper motion and unbiased by any assumptions on spectral energy distributions, might also lead to the discovery of new types of quasars or new classes of extragalactic point sources.




\cleardoublepage

\chapterfont{\raggedright}

\chapter*{Publications}
\addcontentsline{toc}{chapter}{Publications}

Number of refereed first-author articles: 11, citations: 59.\\
Number of refereed articles: 34, citations: 1589.\\
Last updated on: \today. \\

\subsubsection*{Refereed publications}

\begin{enumerate}
	
	\item \textit{Exploring galaxy dark matter haloes across redshifts with strong quasar absorbers}, \\
	Christensen, L., M\o ller, P., Rhodin, N. H. P., \textbf{Heintz, K. E.}, \& Fynbo, J. P. U., Monthly Notices of the Royal Astronomical Society, vol 489, issue 2, 10 pp. (2019)
	
	\item \textit{Short GRB\,160821B: A reverse shock, a refreshed shock, and a well-sampled kilonova}, \\
	Lamb, G. P., Tanvir, N. R., Levan, A. J., de Ugarte Postigo, A., Kawaguchi, K., Corsi, A., Evans, P. A., Gompertz, B., Malesani, D. B., Page, K. L., Wiersema, K., Rosswog, S., Shibata, M., Tanaka, M., van der Horst, A. J., Cano, Z., Fynbo, J. P. U., Fruchter, A. S., Greiner, J., \textbf{Heintz, K. E.}, Higgins, A., Hjorth, J., Izzo, L., Jakobsson, P., Kann, D. A., O’Brien, P. T., Perley, D. A., Pian, E., Pugliese, G., Starling, R. L. C., Th\"one, C. C., Watson, D., Wijers, R. A. M. J., \& Xu, D., The Astrophysical Journal, vol. 883, id. 48, 12 pp. (2019)
	
	\item \textit{New constraints on the physical conditions in H$_2$-bearing GRB-host damped Lyman-$\alpha$ absorbers}, \\
	\textbf{Heintz, K. E.}, Bolmer, J., Ledoux, C., Noterdaeme, P., Krogager, J. -K., Fynbo, J. P. U., Jakobsson, P., Covino, S., D'Elia, V., De Pasquale, M., Hartmann, D. H., Izzo, L., Japelj, J., Kann, D. A., Kaper, L., Petitjean, P., Rossi, A., Salvaterra, R., Schady, P., Selsing, J. ,Starling, R., Tanvir, N. R., Th\"one, C. C., de Ugarte Postigo, A., Vergani, S. D., Watson, D., Wiersema, K., \& Zafar, T., Astronomy \& Astrophysics, vol. 629, id. A131, 21 pp. (2019) 
	
	\item \textit{The effect of dust bias on the census of neutral gas and metals in the high-redshift Universe due to SDSS-II quasar colour selection}, \\
	Krogager, J.-K., Fynbo, J. P. U., M\o ller, P., Noterdaeme, P., \textbf{Heintz, K. E.}, \& Pettini, M., Monthly Notices of the Royal Astronomical Society, vol 486, issue 3, 21 pp. (2019)
	
	\item \textit{On the dust properties of high-redshift molecular clouds and the connection to the 2175\,{\AA} extinction bump}, \\
	\textbf{Heintz, K. E.}, Zafar, T., De Cia, A., Vergani, S. D., Jakobsson, P., Fynbo, J. P. U., Watson, D., Japelj, J., Møller, P., Covino, S., Kaper, L., \& Andersen, A. C., Monthly Notices of the Royal Astronomical Society, vol. 486, issue 2, 12 pp. (2019)
	
	\item \textit{Gaia-assisted selection of a quasar reddened by dust in an extremely strong damped Lyman-$\alpha$ absorber at $z = 2.226$}, \\
	Geier, S. J., \textbf{Heintz, K. E.}, Fynbo, J. P. U., Ledoux, C., Christensen, L., Jakobsson, P., Krogager, J.-K., Milvang-Jensen, B., M\o ller, P., \& Noterdaeme, P., Astronomy \& Astrophysics, vol. 625, id. L9, 5 pp. (2019) 
	
	\item \textit{The fraction of ionizing radiation from massive stars that escapes to the intergalactic medium}, \\
	Tanvir, N. R., Fynbo, J. P. U., de Ugarte Postigo, A., Japelj, J., Wiersema, K., Malesani, D., Perley, D. A., Levan, A. J., Selsing, J., Cenko, S. B., Kann, D. A., Milvang-Jensen, B., Berger, E., Cano, Z., Chornock, R., Covino, S., Cucchiara, A., D'Elia, V., Gargiulo, A., Goldoni, P., Gomboc, A., \textbf{Heintz, K. E.}, Hjorth, J., Izzo, L., Jakobsson, P., Kaper, L., Krühler, T., Laskar, T., Myers, M., Piranomonte, S., Pugliese, G., Rossi, A., S\'anchez-Ram\'irez, R., Schulze, S., Sparre, M., Stanway, E. R., Tagliaferri, G., Thöne, C. C., Vergani, S., Vreeswijk, P. M., Wijers, R. A. M. J., Watson, D., \& Xu, D., Monthly Notices of the Royal Astronomical Society, vol. 483, issue 4, 29 pp. (2019)

	\item \textit{The X-shooter GRB afterglow legacy sample (XS-GRB)}, \\
	Selsing, J., Malesani, D., Goldoni, P., Fynbo, J. P. U., Krühler, T., Antonelli, L. A., Arabsalmani, M., Bolmer, J., Cano, Z., Christensen, L., Covino, S., D'Avanzo, P., D'Elia, V., De Cia, A., de Ugarte Postigo, A., Flores, H., Friis, M., Gomboc, A., Greiner, J., Groot, P., Hammer, F., Hartoog, O. E., \textbf{Heintz, K. E.}, Hjorth, J., Jakobsson, P., Japelj, J., Kann, D. A., Kaper, L., Ledoux, C., Leloudas, G., Levan, A. J., Maiorano, E., Melandri, A., Milvang-Jensen, B., Palazzi, E., Palmerio, J. T., Perley, D. A., Pian, E., Piranomonte, S., Pugliese, G., S{\'a}nchez-Ram{\'{\i}}rez, R., Savaglio, S., Schady, P., Schulze, S., Sollerman, J., Sparre, M., Tagliaferri, G., Tanvir, N. R., Th\"one, C. C., Vergani, S. D., Vreeswijk, P., Watson, D., Wiersema, K., Wijers, R., Xu, D. \& Zafar, T., Astronomy \& Astrophysics, vol. 623, id. A92, 42 pp. (2019)
	
	\item \textit{Evidence for diffuse molecular gas and dust in the hearts of gamma-ray burst host galaxies. Unveiling the nature of high-redshift damped Lyman-$\alpha$ systems}, \\
	Bolmer, J., Ledoux, C., Wiseman, P., De Cia, A., Selsing, J., Schady, P., Greiner, J., Savaglio, S., Burgess, J. M., D'Elia, V., Fynbo, J. P. U., Goldoni, P., Hartmann, D. H., \textbf{Heintz, K. E.}, Jakobsson, P., Japelj, J., Kaper, L., Tanvir, N. R., Vreeswijk, P. M. \& Zafar, T., Astronomy \& Astrophysics, vol. 623, id. A43, 45 pp. (2019)

	\item \textit{The intergalactic magnetic field probed by a giant radio galaxy}, \\
	O'Sullivan, S. P., Machalski, J., Van Eck, C. L., Heald, G., Br\"uggen, M., Fynbo, J. P. U. \textbf{Heintz, K. E.}, Lara-Lopez, M. A., Vacca, V., Hardcastle, M. J., Shimwell, T. W., Tasse, C., Vazza, F., Andernach, H., Birkinshaw, M., Haverkorn, M., Horellou, C., Williams, W. L., Harwood, J. J., Brunetti, G., Anderson, J. M., Mao, S. A., Nikiel-Wroczy\'nski, B., Takahashi, K., Carretti, E., Vernstrom, T., van Weeren, R. J., Orr\'u, E., Morabito, L. K., \& Callingham, J. R., Astronomy \& Astrophysics, vol. 622, id. A16, 12 pp. (2019)
	
	\item \textit{Signatures of a jet cocoon in early spectra of a supernova associated with a $\gamma$-ray burst}, \\
	Izzo, L., de Ugarte Postigo, A., Maeda, K., Th\"one, C. C., Kann, D. A., Della Valle, M., Sagues Carracedo, A., Michałowski, M. J., Schady, P., Schmidl, S., Selsing, J., Starling, R. L. C., Suzuki, A., Bensch, K., Bolmer, J., Campana, S., Cano, Z., Covino, S., Fynbo, J. P. U., Hartmann, D. H., \textbf{Heintz, K. E.}, Hjorth, J., Japelj, J., Kamiński, K., Kaper, L., Kouveliotou, C., Krużyński, M., Kwiatkowski, T., Leloudas, G., Levan, A. J., Malesani, D. B., Michałowski, T., Piranomonte, S., Pugliese, G., Rossi, A., S\'anchez-Ram\'irez, R., Schulze, S., Steeghs, D., Tanvir, N. R., Ulaczyk, K., Vergani, S. D., \& Wiersema, K., Nature, vol. 565, issue 7739, 4 pp. (2019)
	
	\item \textit{Cold gas in the early Universe. Survey for neutral atomic-carbon in GRB host galaxies at $1 < z< 6$ from optical afterglow spectroscopy}, \\
	\textbf{Heintz, K. E.}, Ledoux, C., Fynbo, J. P. U., Jakobsson, P., Noterdaeme, P., Krogager, J.-K., Bolmer, J., Møller, P., Vergani, S. D., Watson, D., Zafar, T., De Cia, A., Tanvir, N. R., Malesani, D. B., Japelj, J., Covino, S., \& Kaper, L., Astronomy \& Astrophysics, vol. 621, id. A20, 13 pp. (2019)
	
	\item \textit{X-shooter and ALMA spectroscopy of GRB 161023A. A study of metals and molecules in the line of sight towards a luminous GRB}, \\
	de Ugarte Postigo, A., Th\"one, C. C., Bolmer, J., Schulze, S., Martín, S., Kann, D. A., D'Elia, V., Selsing, J., Martin-Carrillo, A., Perley, D. A., Kim, S., Izzo, L., S\'anchez-Ram\'irez, R., Guidorzi, C., Klotz, A., Wiersema, K., Bauer, F. E., Bensch, K., Campana, S., Cano, Z., Covino, S., Coward, D., De Cia, A., de Gregorio-Monsalvo, I., De Pasquale, M., Fynbo, J. P. U., Greiner, J., Gomboc, A., Hanlon, L., Hansen, M., Hartmann, D. H., \textbf{Heintz, K. E.}, Jakobsson, P., Kobayashi, S., Malesani, D. B., Martone, R., Meintjes, P. J., Michałowski, M. J., Mundell, C. G., Murphy, D., Oates, S., Salmon, L., van Soelen, B., Tanvir, N. R., Turpin, D., Xu, D., \& Zafar, T., Astronomy \& Astrophysics, vol. 620, id. A119, 23 pp. (2018)
	
	\item \textit{X-shooting GRBs at high redshift: probing dust production history}, \\
	Zafar, T., M\o ller, P., Watson, D., Lattanzio, J., Hopkins, A. M., Karakas, A., Fynbo, J. P. U., Tanvir, N. R., Selsing, J., Jakobsson, P., \textbf{Heintz, K. E.}, Kann, D. A., Groves, B., Kulkarni, V., Covino, S., D'Elia, V., Japelj, J., Corre, D., \& Vergani S.,
	Monthly Notices of the Royal Astronomical Society, vol. 480, issue 1, 10 pp. (2018) 
	
	\item \textit{The properties of GRB\,120923A at a spectroscopic redshift of $z\approx7.8$}, \\ 
	Tanvir, N. R., Laskar, T., Levan, A. J., Perley, D. A., Zabl, J., Fynbo, J. P. U., Rhoads, J., Cenko, S. B., Greiner, J., Wiersema, K., Hjorth, J., Cucchiara, A., Berger, E., Bremer, M. N., Cano, Z., Cobb, B. E., Covino, S., D’Elia, V., Fong, W., Fruchter, A. S., Goldoni, P., Hammer, F., \textbf{Heintz, K. E.}, Jakobsson, P., Kann, D. A., Kaper, L., Klose, S., Knust, F., Krühler, T., Malesani, D., Misra, K., Nicuesa Guelbenzu, A., Pugliese, G., S\'anchez-Ram\'irez, R., Schulze, S., Stanway, E. R., de Ugarte Postigo, A., Watson, D., Wijers, R. A. M. J., \& Xu, D., The Astrophysical Journal, vol. 865, id. 107, 16 pp. (2018)
	
	\item \textit{Highly ionized metals as probes of the circumburst gas in the natal regions of gamma-ray bursts}, \\
	\textbf{Heintz, K. E.}, Watson, D., Jakobsson, P., Fynbo, J. P. U., Bolmer, J., Arabsalmani, M., Cano, Z., Covino, S., D'Elia, V., Gomboc, A., Japelj, J., Kaper, L., Krogager, J.-K., Pugliese, G., S\'anchez-Ram\'irez, R., Selsing, J., Sparre, M., Tanvir, N. R., Th\"one, C. C., de Ugarte Postigo, A., \& Vergani, S. D.,
	Monthly Notices of the Royal Astronomical Society, vol. 479, issue 3, 21 pp. (2018) 
	
	\item \textit{ALMA observations of a metal-rich damped Ly$\alpha$ absorber at $z = 2.5832$: evidence for strong galactic winds in a galaxy group}, \\ 
	Fynbo, J. P. U., \textbf{Heintz, K. E.}, Neeleman, M., Christensen, L.,  Dessauges-Zavadsky, M., Kanekar N., M\o ller, P., Prochaska, J. X., Rhodin, N. H. P., \& Zwaan, M.,
	Monthly Notices of the Royal Astronomical Society, vol. 479, issue 2, 6 pp. (2018) 
	
	\item \textit{VLT/X-shooter GRBs: Individual extinction curves of star-forming regions}, \\
	Zafar, T., Watson, D., M\o ller, P., Selsing, J., Fynbo, J. P. U., Schady, P., Wiersema, K., Levan, A. J., \textbf{Heintz, K. E.}, de Ugarte Postigo, A., D'Elia, V., Jakobsson, P., Bolmer, J., Japelj, J., Covino, S., Gomboc, A., \& Cano, Z.,
	Monthly Notices of the Royal Astronomical Society, vol. 479, issue 2, 12 pp. (2018) 
	
	\item \textit{The host galaxy of the short GRB\,111117A at $z = 2.211$. Impact on the short GRB redshift distribution and progenitor channels}, \\
	Selsing, J., Kr\"uhler, T., Malesani, D., D'Avanzo, P., Schulze, S., Vergani, S. D., Palmerio, J., Japelj, J., Milvang-Jensen, B., Watson, D., Jakobsson, P., Bolmer, J., Cano, Z., Covino, S., D'Elia, V., de Ugarte Postigo, A., Fynbo, J. P. U., Gomboc, A., \textbf{Heintz, K. E.}, Kaper, L., Levan, A. J., Piranomonte, S., Pugliese, G., S\'anchez-Ram\'irez, R., Sparre, M., Tanvir, N. R., Th\"one, C. C., \& Wiersema, K.,
	Astronomy \& Astrophysics, vol. 616, id. A48, 11 pp. (2018) 
	
	\item \textit{Unidentified quasars among stationary objects from Gaia DR2}, \\
	\textbf{Heintz, K. E.}, Fynbo, J. P. U., H\o g, E., M\o ller, P., Krogager, J.-K., Geier, S., Jakobsson, P., \& Christensen, L., 
	Astronomy \& Astrophysics, vol. 615, id. L8, 9 pp. (2018) 
	
	\item \textit{A quasar hiding behind two dusty absorbers. Quantifying the selection bias of metal-rich, damped Ly$\alpha$ absorption systems}, \\
	\textbf{Heintz, K. E.}, Fynbo, J. P. U., Ledoux, C., Jakobsson, P., M\o ller, P., Christensen, L., Geier, S., Krogager, J.-K., \& Noterdaeme, P.,
	Astronomy \& Astrophysics, vol. 615, id. A43, 14 pp. (2018) 
	
	\item \textit{The 2175\,\AA\ Extinction Feature in the Optical Afterglow Spectrum of GRB\,180325A at $z=2.25$}, \\
	Zafar, T., \textbf{Heintz, K. E.}, Fynbo, J. P. U., Malesani, D., Bolmer, J., Ledoux, C., Arabsalmani, M., Kaper, L., Campana, S., Starling, R. L. C., Selsing, J., Kann, D. A., de Ugarte Postigo, A., Schweyer, T., Christensen, L., M\o ller, P., Japelj, J., Perley, D., Tanvir, N. R., D'Avanzo, P., Hartmann, D. H., Hjorth, J., Covino, S., Sbarufatti, B., Jakobsson, P., Izzo, L., Salvaterra, R., D'Elia, V., \& Xu, D.,
	The Astrophysical Journal  Letters, vol. 860, id. L21, 7 pp. (2018) 
	
	\item \textit{The luminous, massive and solar metallicity galaxy hosting the \textit{Swift} $\gamma$-ray burst GRB\,160804A at $z=0.737$}, \\
	\textbf{Heintz, K. E.}, Malesani, D., Wiersema, K., Jakobsson, P., Fynbo, J. P. U., Savaglio, S., Cano, Z., D'Elia, V., Gomboc, A., Hammer, F., Kaper, L., Milvang-Jensen, B., M\o ller, P., Piranomonte, S., Selsing, J., Rhodin, N. H. P., Tanvir, N. R., Th\"one, C. C., de Ugarte Postigo, A., Vergani, S. D., \& Watson, D.,
	Monthly Notices of the Royal Astronomical Society, vol. 474, issue 2, 12 pp. (2018) 
	
	\item \textit{Mass and metallicity scaling relations of high-redshift star-forming galaxies selected by GRBs}, \\
	Arabsalmani, M., M\o ller, P., Perley, D. A., Freudling, W., Fynbo, J. P. U., Le Floc'h, E., Zwaan, M. A., Schulze, S., Tanvir, N. R., Christensen, L., Levan, A. J., Jakobsson, P., Malesani, D., Cano, Z., Covino, S., D'Elia, V., Goldoni, P., Gomboc, A., \textbf{Heintz, K. E.}, Sparre, M., de Ugarte Postigo, A., \& Vergani, S. D.,
	Monthly Notices of the Royal Astronomical Society, vol. 473, issue 3, 13 pp. (2018) 
	
	\item \textit{Solving the conundrum of intervening strong Mg\,\textsc{ii} absorbers towards gamma-ray bursts and quasars}, \\
	Christensen, L., Vergani, S. D., Schulze, S., Annau, N., Selsing, J., Fynbo, J. P. U., de Ugarte Postigo, A., Ca\~nameras, R., Lopez, S., Passi, D., Cort\'es-Zuleta, P., Ellison, S. L., D'Odorico, V., Becker, G., Berg, T. A. M., Cano, Z., Covino, S., Cupani, G., D'Elia, V., Goldoni, P., Gomboc, A., Hammer, F., \textbf{Heintz, K. E.}, Jakobsson, P., Japelj, J., Kaper, L., Malesani, D., M\o ller, P., Petitjean, P., Pugliese, V., S\'anchez-Ram\'irez, R., Tanvir, N. R., Th\"one, C. C., Vestergaard, M., Wiersema, K., \& Worseck, G.,
	Astronomy \& Astrophysics, vol. 608, id. A84, 10 pp. (2017) 
	
	\item \textit{A kilonova as the electromagnetic counterpart to a gravitational-wave source}, \\  
	Smartt, S. J., et al. (2017); co-author \#59 (out of 121), 
	Nature, vol. 551, issue 7678, 51 pp. (2017) 
	
	\item \textit{Multi-messenger Observations of a Binary Neutron Star Merger}, \\
	LIGO/VIRGO \& follow-up collaboration; co-author as part of ePESSTO,
	The Astrophysical Journal  Letters, vol. 848, id. L12, 59 pp. (2017) 
	
	\item \textit{The High $A_V$ Quasar survey: A $z=2.027$ metal-rich damped Lyman-$\alpha$ absorber towards a red quasar at $z=3.21$}, \\
	Fynbo, J. P. U., Krogager, J.-K., \textbf{Heintz, K. E.}, Geier, S., M\o ller, P., Noterdaeme, P., Christensen, L., Ledoux, C., \& Jakobsson, P.,
	Astronomy \& Astrophysics, vol. 606, id. A13, 6 pp. (2017) 
	
	\item \textit{GRB\,161219B / SN\,2016jca: A low-redshift gamma-ray burst supernova powered by radioactive heating}, \\ 
	Cano, Z., Izzo, L., de Ugarte Postigo, A., Th\"one, C. C., Kr\"uhler, T., \textbf{Heintz, K. E.}, Malesani, D., Geier, S., Fuentes, C., Chen, T.-W., Covino, S., D'Elia, V., Fynbo, J. P. U., Goldoni, P., Gomboc, A., Hjorth, J., Jakobsson, P., Kann, D. A., Milvang-Jensen, B., Pugliese, G., S\'anchez-Ram\'irez, R., Schulze, S., Sollerman, J., Tanvir, N. R., \& Wiersema, K., 
	Astronomy \& Astrophysics, vol. 605, id. A107, 21 pp. (2017) 
	
	\item \textit{Steep extinction towards GRB\,140506A reconciled from host galaxy observations: Evidence that steep reddening laws are local}, \\ 
	\textbf{Heintz, K. E.}, Fynbo, J. P. U., Jakobsson, P., Kr\"{u}hler, T., Christensen, L., Watson, D., Ledoux, C., Noterdaeme, P., Perley, D. A., Rhodin, H., Selsing, J., Schulze, S., Tanvir, N. R., M\o ller, P., Goldoni, P., Xu, D., \& Milvang-Jensen, B.,
	Astronomy \& Astrophysics, vol. 601, id. A83, 10 pp. (2017) 
	
	\item \textit{The Extended High A(V) Quasar Survey: Searching for Dusty Absorbers Toward Mid-infrared Selected Quasars}, \\ 
	Krogager, J.-K., Fynbo, J. P. U., \textbf{Heintz, K. E.}, Geier, S., Ledoux, C., M\o ller, P., Noterdaeme, P., Venemans, B., \& Vestergaard, M.,
	The Astrophysical Journal, vol. 832, id. 49, 21 pp. (2016) 
	
	\item \textit{Determining the fraction of reddened quasars in COSMOS with multiple selection techniques from X-ray to radio wavelengths}, \\ 
	\textbf{Heintz, K. E.}, Fynbo, J. P. U., M\o ller, P., Milvang-Jensen, B., Zabl, J., Maddox, N., Krogager, J.-K., Geier, S., Vestergaard, M., Noterdaeme, P., \& Ledoux, C.,
	Astronomy \& Astrophysics, vol. 595, id. A13, 22 pp. (2016)
	
	\item \textit{Serendipitous Discovery of a Projected Pair of QSOs Separated by 4.5 arcsec on the Sky}, \\ 
	\textbf{Heintz, K. E.}, Fynbo, J. P. U., Krogager, J.-K., Vestergaard, M., M\o ller, P., Arabsalmani, M., Geier, S., Noterdaeme, P., Ledoux, C., Saturni, F. G., \& Venemans, B.,
	The Astronomical Journal, vol. 152, id. 13, 4 pp. (2016) 
	
	\item \textit{A study of purely astrometric selection of extragalactic point sources with Gaia}, \\ 
	\textbf{Heintz, K. E}., Fynbo, J. P. U., \& H\o g, E.,
	Astronomy \& Astrophysics, vol. 578, id. A91, 4 pp. (2015)
\end{enumerate}
%
%
%
\subsubsection*{GRB coordinate network circulars (GCNs)}

\begin{enumerate}
	\setlength\itemsep{0.01em}
	\item \textit{GRB\,191004B: VLT/X-shooter redshift}, \\
	D'Elia, V., Fynbo, J. P. U., Izzo, L., Malesani, D. B., \textbf{Heintz, K. E.}, Tanvir, N. R., de Ugarte Postigo, A., and Vergani, S. D., GCN circular \#~25956 (2019).
	\item \textit{GRB\,191004B: NOT optical photometry}, \\
	\textbf{Heintz, K. E.}, Malesani, D. B., and Moran, S., GCN circular \#~25954 (2019).
	\item \textit{GRB\,191004A: NOT optical upper limits}, \\
	\textbf{Heintz, K. E.}, Malesani, D. B., and Moran, S., GCN circular \#~25952 (2019).
	\item \textit{GRB\,190829A: NOT optical afterglow detection and spectroscopy}, \\
	\textbf{Heintz, K. E.}, Fynbo, J. P. U., Jakobsson, P., Xu, D., Perley, D. A., Malesani, D. B., and Viuho, J., GCN circular \#~25563 (2019)
	\item \textit{GRB\,190719C: VLT/X-shooter spectroscopic redshift of the host galaxy}, \\
	Rossi, A., \textbf{Heintz, K. E.}, Fynbo, J. P. U., Malesani, D. B., de Ugarte Postigo, A., Vergani, S. D., Kann, D. A., Thoene, C. C., Izzo, L., Campana, S., Pugliese, G., Kaper, L., Kouveliotou, C., and Tanvir, N. R., GCN circular \#~25252 (2019).
	\item \textit{GRB\,190719C: NOT optical afterglow confirmation}, \\
	\textbf{Heintz, K. E.}, Malesani, D. B., Perley, D. A., and Moran, S., GCN circular \#~25144 (2019).
	\item \textit{GRB\,190719C: NOT candidate optical afterglow and host}, \\
	Malesani, D. B., \textbf{Heintz, K. E.}, Perley, D. A., Milvang-Jensen, B., and Moran, S., GCN circular \#~25110 (2019).
	\item \textit{GRB\,190613B: Optical observations from NOT}, \\
	de Ugarte Postigo, A., Malesani, D. B.,  \textbf{Heintz, K. E.}, and Moran, S., GCN circular \#~24825 (2019).
	\item \textit{GRB\,190531B: VLT optical upper limit}, \\
	Japelj, J., Rossi, A., Malesani, D. B., \textbf{Heintz, K. E.}, Pugliese, G., and Kaper, L., GCN circular \#~24711 (2019).
	\item \textit{GRB\,190530A: NOT photometry and spectroscopy}, \\
	\textbf{Heintz, K. E.}, Fynbo, J. P. U., de Ugarte Postigo, A., Malesani, D. B., Selsing, J., Milvang-Jensen, B., and Moran, S., GCN circular \#~24686 (2019).
	\item \textit{LIGO/Virgo S190408an: GROND Observations of MASTER OT J154209.55-431742.2}, \\
	Chen, T.-W., Schweyer, T., Rossi, A., \textbf{Heintz, K. E.}, Gromadzki, M., Bolmer, J., and Schady, P., GCN circular \#~24097 (2019).
	\item \textit{GRB\,190219A: NOT optical afterglow candidate}, \\
	Xu, D., \textbf{Heintz, K. E.}, Malesani, D. B., Zhu, Z. P., Galindo, P., and Viuho, J., GCN circular \#~23911 (2019).
	\item \textit{GRB\,190211A: NOT optical observations}, \\
	\textbf{Heintz, K. E.}, Xu, D., Malesani, D. B., and Moran, S., GCN circular \#~23890 (2019).
	\item \textit{GRB\,190114C: X-shooter observations of a highly extinguished afterglow}, \\
	Kann, D. A., Th\"one, C. C., Selsing, J., Izzo, L., de Ugarte Postigo, A., Pugliese, G., Sbarufatti, B., \textbf{Heintz, K. E.}, D'Elia, V., Covino, S., Wiersema, K., Perley, D. A., Vergani, S., Fynbo, J. P. U., Watson, D., Tanvir, N. R., Hartmann, D., Xu, D., Schulze, S., and Bolmer, J., GCN circular \#~23710 (2019).
	\item \textit{GRB\,190114A: NOT optical observations}, \\
	Selsing, J., de Ugarte Postigo, A., and \textbf{Heintz, K. E.}, GCN circular \#~23697 (2019).
	\item \textit{GRB\,190114C: NOT optical counterpart and redshift}, \\
	Selsing, J., Fynbo, J. P. U., \textbf{Heintz, K. E.}, and Watson, D., GCN circular \#~23695 (2019).
	\item \textit{GRB 190106A: VLT/X-shooter redshift confirmation}, \\
	Schady, P., Xu, D., \textbf{Heintz, K. E.}, Tanvir, N. R., Malesani D. B., Kann, D. A., S\'anchez-Ram\'irez, R., and Wiersema, K., GCN circular \#~23632 (2019).
	\item \textit{GRB\,181213A: NOT optical observations}, \\
	\textbf{Heintz, K. E.}, Malesani, D. B., Fynbo, J. P. U., de Ugarte Postigo, A., Balaguer-Nu$\tilde{\rm n}$ez, L., Carbajo, J., Galindo, F., and Perez, C., GCN circular \#~23537 (2018).
	\item \textit{GRB\,181201A: VLT/FORS2 tentative spectroscopic redshift}, \\
	Izzo, L., de Ugarte Postigo, A., Kann, D. A., Malesani, D. B., \textbf{Heintz, K. E.}, Tanvir, N. R., D'Elia, V., Wiersema, K., Kouveliotou, C., and Levan, A. J., GCN circular \#~23488 (2018).
	\item \textit{GRB\,181201A: NOT optical observations}, \\
	\textbf{Heintz, K. E.}, Malesani, D. B., and Moran-Kelly, S., GCN circular \#~23478 (2018).
	\item \textit{GRB\,181110A: VLT/X-shooter redshift}, \\
	Perley, D. A., Malesani, D. B., Fynbo, J. P. U., \textbf{Heintz, K. E.}, Kann, D. A., D'Elia, V., Izzo, L., and Tanvir, N. R., GCN circular \#~23421 (2018).
	\item \textit{GRB\,181020A: VLT/X-shooter spectroscopy and redshift}, \\
	Fynbo, J. P. U., de Ugarte Postigo, A., D'Elia, V., Tanvir, N. R., Pugliese, G., Kann, D. A., Malesani, D. B., and \textbf{Heintz, K. E.}, GCN circular \#~23356 (2018).
	\item \textit{GRB\,181010A: VLT/X-shooter redshift}, \\
	Vielfaure, J.-B., Japelj, J., Malesani, D. B., D'Elia, V., Fynbo, J. P. U., Tanvir, N. R., Vergani, S. D., Rossi, A., Vreeswijk, P., Izzo, L., Kann, D. A., Milvang-Jensen, B., Xu, D., and \textbf{Heintz, K. E.}, GCN circular \#~23315 (2018).
	\item \textit{GRB\,180914B: VLT/X-shooter spectroscopic redshift}, \\
	D'Avanzo, P., \textbf{Heintz, K. E.}, de Ugarte Postigo, A., Levan, A. J., Izzo, L., Malesani, D. B., Kann, D. A., and Tanvir, N. R., GCN circular \#~23246 (2018).
	\item \textit{GRB\,180904A: NOT optical observations, candidate host galaxy}, \\
	Malesani, D., \textbf{Heintz, K. E.}, and Dyrbye, S., GCN circular \#~23195 (2018).
	\item \textit{GRB\,180728A: Classification of the associated SN 2018fip}, \\
	Selsing, J., Izzo, L., Rossi, A., Malesani, D. B., \textbf{Heintz, K. E.}, de Ugarte Postigo, A., Schady, P., Starling, R. L. C., Sollerman, J., Leloudas, G., Cano, Z., Fynbo, J. P. U., Della Valle, M., Pian, E., Kann, D. A., Perley, D. A., Palazzi, E., Klose, S., Hjorth, J., Covino, S., D'Elia, V., Tanvir, N. R., Levan, A. J., Hartmann, D., and Kouveliotou C., GCN circular \#~23181 (2018).
	\item \textit{GRB\,180728A: Discovery of the associated supernova}, \\	Izzo, L., Rossi, A., Malesani, D. B., \textbf{Heintz, K. E.}, Selsing, J., Schady, P., Starling, R. L. C., Sollerman, J., Leloudas, G., Cano, Z., Fynbo, J. P. U., Della Valle, M., Pian, E., Kann, D. A., Perley, D. A., Palazzi, E., Klose, S., Hjorth, J., Covino, S., D'Elia, V., Tanvir, N. R., Levan, A. J., Hartmann, D., and Kouveliotou C., GCN circular \#~23142 (2018).
	\item \textit{GRB 180809B: VLT optical upper limits}, \\
	Japelj, J., Malesani, D. B., Selsing, J., Tanvir, N. R., Kann, D. A., D'Elia, V., Pugliese, G., \textbf{Heintz, K. E.}, Fynbo, J. P. U., Levan, A. J., and Schady, P., GCN circular \#~23114 (2018).
	\item \textit{GRB\,180728A: No evidence of SN in early VLT/X-shooter spectra}, \\
	\textbf{Heintz, K. E.}, Izzo, L., Rossi, A., de Ugarte Postigo, A., Malesani, D. B., Perley, D. A., Th\"one, C. C., Fynbo, J. P. U., Milvang-Jensen, B., Kann, D. A., Tanvir, N. R., Levan, A. J., Schulze, S., and Pugliese, G., GCN circular \#~23067 (2018).
	\item \textit{GRB\,180720B: VLT/X-shooter redshift}, \\
	Vreeswijk, P. M., Kann, D. A., \textbf{Heintz, K. E.}, de Ugarte Postigo, A., Milvang-Jensen, B., Malesani, D. B., Covino, S., Levan, A. J., and Pugliese, G., GCN circular \#~22996 (2018).
	\item \textit{GRB\,180624A: NOT optical counterpart}, \\
	Selsing, J., \textbf{Heintz, K. E.}, Fynbo, J. P. U., Malesani, D., Lehtinen, J. J., and Willamo, T., GCN circular \#~22867 (2018).
	\item \textit{GRB\,180620B: VLT/X-shooter redshift}, \\ Izzo, L., Arabsalmani, M., Malesani, D. B., Wiersema, K., Pugliese, G., Vergani, S. D., de Ugarte Postigo, A., \textbf{Heintz, K. E.}, Kann, D. A., Schady, P., Tanvir, N. R., Fynbo, J. P. U., Th\"one, C. C., and Bolmer, J., GCN circular \#~22823 (2018). 
	\item \textit{GRB\,180614A: NOT optical upper limits}, \\
	\textbf{Heintz, K. E.}, Fynbo, J. P. U., Malesani, D., Lehtinen, J. J., Willamo, T., and Galindo, P., GCN circular \#~22784 (2018).
	\item \textit{GRB\,180512A: NOT upper limits}, \\
	\textbf{Heintz, K. E.}, Malesani, D., Siltala, L., and Rodrigues, P. M., GCN circular \#~22713 (2018).
	\item \textit{GRB\,180418A: NOT optical observations}, \\
	Malesani, D., \textbf{Heintz, K. E.}, Stone, M., and Stone, J., GCN circular \#~22660 (2018).
	\item \textit{GRB\,180329B: NOT optical afterglow detection}, \\
	Perley, D. A., \textbf{Heintz, K. E.}, and Malesani, D., GCN circular \#~22562 (2018).
	\item \textit{GRB\,180325A: VLT/X-shooter spectroscopic observations}, \\
	D'Avanzo, P., Bolmer, J., D'Elia, V., Fynbo, J. P. U., \textbf{Heintz, K. E.}, Izzo, L., Japelj, J., Kann, D. A., Malesani, D., Selsing, J., Tanvir, N. R., Zafar, T., Campana, S., de Ugarte Postigo, A., Hjorth, J., Kaper, L., Melandri, A., and Smette, A., GCN circular \#~22555 (2018).
	\item \textit{GRB\,180325A: NOT redshift}, \\
	\textbf{Heintz, K. E.}, Fynbo, J. P. U., and Malesani, D., GCN circular \#~22535 (2018).
	\item \textit{GRB\,180314A: VLT/X-shooter redshift}, \\ Sbarufatti, B., Bolmer, J., de Ugarte Postigo, A., Fynbo, J. P. U., Selsing, J., \textbf{Heintz, K. E.}, Malesani, D., Tanvir, N. R., Levan, A. J., Smette, A., Wiersema, K., and Covino, S., GCN circular \#~22484 (2018).
	\item \textit{GRB\,180205A: VLT/X-shooter redshift}, \\ Tanvir, N. R., \textbf{Heintz, K. E.}, Selsing, J., Japelj, J., Bolmer, J., Kann, D. A., Xu, D., de Ugarte Postigo, A., Malesani, D., Fynbo, J. P. U., and Pugliese, G., GCN circular \#~22384 (2018).
	\item \textit{GRB\,180115A: NOT detection of optical counterpart}, \\ Cano, Z., \textbf{Heintz, K. E.}, de Ugarte Postigo, A., Marquez, I., and Cazzoli, S., GCN circular \#~22345 (2018).
	\item \textit{GRB\,180102A: NOT optical observations}, \\ Malesani, D., \textbf{Heintz, K. E.}, Sagues Carracedo, A., and Dideriksen, A. K., GCN circular \#~22298 (2018).
	\item \textit{GRB\,171222A: Redshift from 10.4m GTC/OSIRIS}, \\ de Ugarte Postigo, A.,Izzo, L., Kann, D. A., Th\"one, C. C., Perley, D. A., Malesani, D., \textbf{Heintz, K. E.}, and Garcia, D., GCN circular \#~22272 (2017).
	\item \textit{GRB\,171222A: NOT optical afterglow}, \\ Perley, D. A., Malesani, D., \textbf{Heintz, K. E.}, de Ugarte Postigo, A., and Lehiten, J. J., GCN circular \#~22269 (2017).
	\item \textit{GRB\,171010A: VLT spectroscopic identification of the associated SN 2017htp}, \\ de Ugarte Postigo, A., Selsing, J., Malesani, D., Xu, D., Izzo, L., \textbf{Heintz, K. E.}, Kann, D. A., Leloudas, G., Schulze, S., Tanvir, N. R., Covino, S., D'Avanzo, P., Fynbo, J. P. U., Hartmann, D., Hjorth, J., Kouveliotou, C., Kaper, L., Levan, A. J., Melandri, A., M\o ller, P., Pugliese, G., Sbarufatti, B., Schady, P., and Schmidl, S., GCN circular \#~22096 (2017).
	\item \textit{GRB\,171027A: VLT optical upper limits}, \\ Malesani, D., Tanvir, N. R., Schady, P., de Ugarte Postigo, A., \textbf{Heintz, K. E.}, and Pugliese, G., GCN circular \#~22060 (2017).
	\item \textit{GRB\,171020A: NOT spectroscopic redshift}, \\ Malesani, D., Fynbo, J. P. U.,  \textbf{Heintz, K. E.}, Stone, M., and Karhunen, K., GCN circular \#~22039 (2017).
	\item \textit{GRB\,171010A: ePESSTO NTT spectroscopic redshift}, \\ Kankare, E., O'Neill, D., Izzo, L., D'Elia, V., Vergani, S. D., Malesani, D., \textbf{Heintz, K. E.}, Schady, P., Melandri, A., D'Avanzo, P., Campana, S., Covino, S., Magee, M., Chen, T.-W., Galbany, L., Inserra, C., Maguire, K., Smartt, S., Yaron, O., Young, D., and Manulis, I., GCN circular \#~22002 (2017).
	\item \textit{GRB\,171010A: NOT optical observations}, \\ Malesani, D., \textbf{Heintz, K. E.}, Calissendorff, P., and Losada, I. R., GCN circular \#~22000 (2017).
	\item \textit{GRB\,170921A: NOT optical observations}, \\ Malesani, D., \textbf{Heintz, K. E.}, Tanvir, N. R., Kann, D. A., Pursimo, T., and Dyrbye, S., GCN circular \#~21909 (2017).
	\item \textit{GRB\,170903A: Host galaxy redshift from OSIRIS/GTC}, \\ de Ugarte Postigo, A., Izzo, L., Th\"one, C. C., Kann, D. A., \textbf{Heintz, K. E.}, Castro-Rodrigues, N., and Marante, A., GCN circular \#~21799 (2017).
	\item \textit{GRB\,170728B: NOT optical afterglow confirmation}, \\ \textbf{Heintz, K. E.}, Tanvir, N. R., Perley, D A., and Rasmussen, R. T., GCN circular \#~21374 (2017)
	\item \textit{GRB\,170607A: NOT optical afterglow detection}, \\ \textbf{Heintz, K. E.}, de Ugarte Postigo, A., and Claesen, J.,  GCN circular \#~21217 (2017) 
	\item \textit{GRB\,170519A: NOT optical observations}, \\ de Ugarte Postigo, A., Xu, D., Malesani, D., \textbf{Heintz, K. E.}, Gandolfi, D., and Telting, J., GCN circular \#~21120 (2017) 
	\item \textit{GRB\,170405A: NOT optical observations}, \\ Malesani, D., \textbf{Heintz, K. E.} and Pursimo, T., GCN circular \#~20988 (2017) 
	\item \textit{GRB\,170205A: NOT upper limits}, \\ \textbf{Heintz, K. E.}, Malesani, D., Tanvir, N. R., de Ugarte Postigo, A., Korn, A. J. and Losada, I. R., GCN circular \#~20638 (2017)
	\item \textit{GRB\,170205A: NOT afterglow observations}, \\ Kr\"{u}hler, T., Xu, D., \textbf{Heintz, K. E.} and Fedorets, G., GCN circular \#~20612 (2017)
	\item \textit{GRB\,170127B: NOT r-band upper limits}, \\ Cano, Z., \textbf{Heintz, K. E.}, Malesani, D., and Tronsgaard, R., GCN circular \#~20549 (2017)
	\item \textit{GRB\,170113A: VLT/X-shooter redshift}, \\ Xu, D., \textbf{Heintz, K. E.}, Malesani, D., and Fynbo, J. P. U., GCN circular \#~20458 (2017)
	\item \textit{GRB 161129A: NOT optical observations and candidate host galaxy}, \\ \textbf{Heintz, K. E.}, Malesani, D., Cano, Z., de Ugarte Postigo, A., Fedorets, G., Tronsgaard Rasmussen, R., Djupvik, A. A., Losada, I. R., Svardh, I., Clasen, J. and Armas Perez, S., GCN circular \#~20244 (2016)
	\item \textit{GRB 161117A: VLT/X-shooter spectroscopy and redshift}, \\ Malesani, D., Kr\"{u}hler, T., \textbf{Heintz, K. E.} and Fynbo, J. P. U., GCN circular \#~20180 (2016)
	\item \textit{GRB\,161108A: NOT redshift}, \\ de Ugarte Postigo, A., Malesani, D., Perley, D. A., Fynbo, J. P. U., \textbf{Heintz, K. E.}, Somero, A., Gafton, E., Damsted, S., Erfanianfar, G., Finoguenov, A., Gibson, C., Kiefer, F., Kirkpatrick, C., Lumme, M., Oja, V., Rantakyla, J., Salmenpera, I. and Seppala, M., GCN circular \#~20150 (2016)
	\item \textit{GRB\,161014A: VLT/X-shooter redshift confirmation}, \\ Selsing, J., \textbf{Heintz, K. E.}, Malesani, D., Xu, D., de Ugarte Postigo, A., Tanvir, N. R., Wiersema, K. and Fynbo, J. P. U., GCN circular \#~20061 (2016)
	\item \textit{GRB\,161014A: NOT optical observations}, \\ \textbf{Heintz, K. E.}, Xu, D., Malesani, D., Poretti, E., Telting, J., de Ugarte Postigo, A. and Fynbo, J. P. U., GCN circular \#~20044 (2016)
	\item \textit{GRB\,161007A: NOT optical observations}, \\ \textbf{Heintz, K. E.}, Malesani, D., de Ugarte Postigo, A., Pursimo, T., Kruehler, T., Telting, J. and Fynbo, J. P. U., GCN circular \#~20020 (2016)
	\item \textit{GRB\,160804A: VLT/X-shooter redshift}, \\ Xu, D., \textbf{Heintz, K. E.}, Malesani, D., Wiersema, K., and Fynbo, J. P. U., GCN circular \#~19773 (2016)
	\item \textit{GRB\,150314A: NOT optical observations}, \\ Xu, D., de Ugarte Postigo, A., Geier, S., \textbf{Heintz, K. E.}, Fynbo, J. P. U., Malesani, D., and Jakobsson, P., GCN circular \#~17582 (2015)
\end{enumerate}


\subsubsection*{Astronomer's telegrams (ATels)}

\begin{enumerate}
	\setlength\itemsep{0.05em}
	\item \textit{GRB\,180728A: No evidence of SN in early VLT/X-shooter spectra}, \\
	\textbf{Heintz, K. E.}, Izzo, L., Rossi, A., de Ugarte Postigo, A., Malesani, D. B., Perley, D. A., Th\"one, C. C., Fynbo, J. P. U., Milvang-Jensen, B., Kann, D. A., Tanvir, N. R., Levan, A. J., Schulze, S., and Pugliese, G., ATel \#~11898 (2018).
	\item \textit{ePESSTO spectroscopic classification of optical transients}, \\
	Malesani, D., \textbf{Heintz, K. E.}, Leloudas, G., D'Avanzo, P., Barbarino, C., Chen, T. W., Kostrzewa-Rutkowska, Z., Congiu, E., Gromadzki, M., Inserra, C., Kankare, E., Maguire, K., Smartt, S. J., Yaron, O., Young, D., Manulis, I., Tonry, J., Denneau, L., Heinze, A., Weiland, H., Stalder, B., Rest, A., Smith, K. W., McBrien, O., Wright, D. E., Chambers, K. C., Flewelling, H., Huber, M., Lowe, T., Magnier, E., Schultz, A., Waters, C., Wainscoat, R. J., and Wilman, M., ATel \#~11564 (2018).
	\item \textit{ePESSTO spectroscopic classification of optical transients}, \\
	Malesani, D., \textbf{Heintz, K. E.}, Leloudas, G., D'Avanzo, P., Barbarino, C., Chen, T. W., Kostrzewa-Rutkowska, Z., Gromadzki, M., Onori, F., Inserra, C., Kankare, E., Maguire, K., Smartt, S. J., Yaron, O., Young, D., Smith, K. W., Wright, D. E., Chambers, K. C., Flewelling, H., Huber, M., Lowe, T., Magnier, E., Schultz, A., Waters, C., Wainscoat, R. J., and Wilman, M., ATel \#~11560 (2018).
	\item \textit{ePESSTO spectroscopic classification of optical transients}, \\ \textbf{Heintz, K. E.}, Malesani, D., D'Avanzo, P., Leloudas, G., Barbarino, C., Chen, T. W., Kostrzewa-Rutkowska, Z., Inserra, C., Kankare, E., Maguire, K., Smartt, S. J., Yaron, O., Young, D., and Manulis, I., ATel \#~11556 (2018).
	\item \textit{ePESSTO spectroscopic classification of optical transients}, \\ \textbf{Heintz, K. E.}, Malesani, D., D'Avanzo, P., Leloudas, G., Barbarino, C., Chen, T. W., Kostrzewa-Rutkowska, Z., Inserra, C., Kankare, E., Maguire, K., Smartt, S. J., Yaron, O., Young, D., Tonry, J., Denneau, L., Heinze, A., Weiland, H., Stalder, B., Rest, A., Smith, K. W., McBrien, O., Young, D. R., Wright, D. E., and Wyrzykowski, L., ATel \#~11554 (2018).
	\item \textit{ePESSTO spectroscopic classification of optical transients}, \\ Malesani, D., Rubin, A., Leloudas, G., \textbf{Heintz, K. E.}, Dennefeld, M., Galbany, L., Benetti, S., Taubenberger, S., Inserra, C., Kankare, E., Maguire, K., Smartt, S. J., Yaron, O., and Young, D., ATel \#~11524 (2018).
	\item \textit{ePESSTO spectroscopic classification of optical transients}, \\ Rubin, A., Malesani, D., \textbf{Heintz, K. E.}, Dennefeld, M., Galbany, L., Gromadzki, M., Benetti, S., Inserra, C., Kankare, E., Maguire, K., Smartt, S. J., Yaron, O., and Young, D., ATel \#~11519 (2018).
	\item \textit{ePESSTO spectroscopic classification of optical transients}, \\ Malesani, D., Rubin, A., Leloudas, G., \textbf{Heintz, K. E.}, Anderson, J., Maguire, K., Inserra, C., Kankare, E., Smartt, S. J., Yaron, O., Young, D., Tonry, J., Denneau, L., Heinze, A., Weiland, H., Stalder, B., Rest, A., Smith, K. W., McBrien, O., Wright, D. E., Chambers, K. C., Flewelling, H., Huber, M., Lowe, T., Magnier, E., Schultz, A., Waters, C., Wainscoat, R. J., and Wilman, M., ATel \#~11509 (2018).
\end{enumerate}

\cleardoublepage

\bibliography{thesisref}

\end{document}